\documentclass[10pt,conference,letterpaper]{IEEEtran}
\usepackage{times,amsmath,epsfig,xcolor}
\usepackage{subfigure}
\usepackage{algorithmic}
\usepackage[ruled,vlined,linesnumbered]{algorithm2e}
\usepackage{color}
\newcommand{\eat}[1]{}

\newtheorem{definition}{Definition}[section]

\newtheorem{theorem}{Theorem}[section]

\title{Trade-offs Computing Minimum Hub Cover toward Optimized Graph Query Processing}
\author{%
{Belma Yelbay{\small $~^{\flat 1}$}, \c{S}. \.{I}lker Birbil{\small $~^{\flat 2}$}, Kerem
B\"{u}lb\"{u}l{\small $~^{\flat 3}$},
Hasan M. Jamil{\small $~^{\natural *}$} }%
\vspace{1.6mm}\\
\fontsize{10}{10}\selectfont\itshape $~^{\flat}$Manufacturing
Systems and Industrial Engineering Department, Sabanci University,
Istanbul, Turkey\\
\fontsize{9}{9}\selectfont\ttfamily\upshape
$~^{1}$byelbay@sabanciuniv.edu, $~^{2}$sibirbil@sabanciuniv.edu,
$~^{3}$bulbul@sabanciuniv.edu
\vspace{1.2mm}\\
\fontsize{10}{10}\selectfont\rmfamily\itshape
$~^{\natural}$Department of Computer Science, University of Idaho,
Moscow, Idaho, USA\\
\fontsize{9}{9}\selectfont\ttfamily\upshape
jamil@uidaho.edu\\
\fontsize{9}{9}\selectfont\rmfamily\itshape $~^{*}$Author for
correspondence }
\begin{document}
\maketitle
\begin{abstract}

As techniques for graph query processing mature, the need for
optimization is increasingly becoming an imperative. Indices are one
of the key ingredients toward efficient query processing strategies
via cost-based optimization. Due to the apparent absence of a common
representation model, it is difficult to make a focused effort
toward developing access structures, metrics to evaluate query
costs, and choose alternatives. In this context, recent interests in
covering-based graph matching appears to be a promising direction of
research. In this paper, our goal is to formally introduce a new
graph representation model, called {\em Minimum Hub Cover}, and
demonstrate that this representation offers interesting strategic
advantages, facilitates construction of candidate graphs from graph
fragments, and helps leverage indices in novel ways for query
optimization. However, similar to other covering problems, minimum
hub cover is NP-hard, and thus is a natural candidate for
optimization. We claim that computing the minimum hub cover leads to
substantial cost reduction for graph query processing. We present a
computational characterization of minimum hub cover based on integer
programming to substantiate our claim and investigate its
computational cost on various graph types.

\end{abstract}

\eat{
\begin{keywords}
ignore
\end{keywords}
}

\section{Introduction}

Queries over graph databases can be classified broadly into whole
graph at-a-time, and node at-a-time processing, and framed as a
subgraph isomorph computation problem (e.g., \cite{KouLM10,YuanM11})
under a set of label mapping constraints, generally known as graph
matching. Techniques such as GraphQL \cite{GraphQL}, QuickSI
\cite{Shang2008} and earlier research such as VFLib \cite{vflib} and
Ullmann \cite{Ullman1976} fall in the former category while TALE
\cite{tales}, and SAPPER \cite{ZhangYJ10} are representative of the
latter. The advantage of the node at-a-time graph processing
approach is its inherent ability to prune search space based on
target node matching conditions. Node indices are the most common
pruning aid used in most of these processing methods although
indices on paths \cite{GraphGrep2002}, frequent structures
\cite{GADDI}, node distances \cite{KouLM10,GADDI,ZagerV08}, etc.
have also been used. The key difference is in the ways indices are
exploited for the construction of the target database graphs from
their parts (i.e., the edges).

The effectiveness of the node at-a-time methods, however, largely
depends on the query type such as subgraph isomorphism, approximate
matching, path queries and so on, as well as on the index structure
used. In other words, universal indexing methods are not always
suitable for all queries, and therefore specialized indices are
often constructed to process a query (e.g., GraphGrep
\cite{GraphGrep2002}, TALE, and SAPPER) and never maintained. Thus,
it is not apparent if the index structure is switched, how an
algorithm will perform leaving an open question if generic index
structures can be leveraged in a way similar to relational query
processing with popular indices such as B+ trees, extendible hashing
and inverted files.

\begin{figure}[h]
\begin{center}
\resizebox{.49\textwidth}{1.25in}{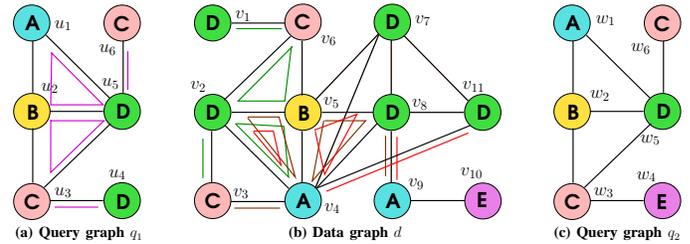}
\caption{Example: Query graphs $q_1$ and $q_2$, and data graph $d$.}
\label{fig:example}
\end{center}
\end{figure}

In order to decouple the index selection from the query expressions,
and to subsequently use indices as a strategic instrument to compute
alternative query plans, we focus on a representation method for
graphs that is independent of the underlying access structures. Our
goal is to propose the ``hub" as the unit of graph representation
that tells us all we need to know about a node or vertex of a graph.
Intuitively, each node in a graph as hub ``covers" all the edges
involving its neighbors and itself. For example, the hub $u_5$ in
figure \ref{fig:example}(a) covers the edges $(u_1,u_5), (u_1,u_2),
(u_2,u_5), (u_2,u_3), (u_3,u_5)$ and $(u_5,u_6)$ as a unit structure
(shown as the purple edges).

The concept of hub we have in mind can be thought of as a convenient
extension of Ullmann's adjacency matrix \cite{Ullman1976} and
feature structure indexing \cite{KardesG10,ChenYYHZG07} in that we
localize the adjacency matrix at the node level and consider only a
single feature, edges among the neighbors. Consequently only
structures that are part of a hub are stars (neighbors with no
shared edge with other neighbors) and triangles (neighbors sharing
edge with other neighbors). In figure \ref{fig:example}(a), vertex
$u_5$ has two triangles $\triangle u_1u_2u_5$ and $\triangle
u_2u_3u_5$, and a star $(u_5,u_6)$ (in this case just an edge).
Whereas vertices $v_9$ and $v_{11}$ in figure \ref{fig:example}(b)
have a star (with two edges $(v_8,v_9)$ and $(v_9,v_{10})$ with
$v_9$ as their center), and three triangles ($\triangle
v_4v_7v_{11}$, $\triangle v_8v_7v_{11}$, and $\triangle
v_4v_8v_{11}$) respectively.

Our goal is to use these atomic structural cues to match shapes for
the purpose of graph matching. For example, to match the query graph
$q_1$ in figure \ref{fig:example}(a) with the data graph $d$ in
figure \ref{fig:example}(b), we look for individual node structures
that are identically connected and depending on the matching
requirement, have identical labels. The next step is to piece
together these individual matches to see if the composed structure
is the target graph. The cost of this search usually is dominated by
the cost of piecing together the components and testing if the
process is yielding the target graph.

In this example, we can contemplate several different types of graph
matching that can be conceived as the variants of subgraph
isomorphism though in the literature, only the structural isomorphs
and match isomorphs defined below are prevalent. We therefore will
consider only these two types of matching in the remainder of this
paper.
\begin{itemize}
\item {\em structural subgraph isomorph}, where only the node IDs (not the labels) are mapped from query graphs to data graphs using and injective function.
\item {\em label subgraph isomorph}, on the other hand, requires an injective mapping of both node IDs and node labels from query graph to data graph.
\item {\em full subgraph isomorph} extends label subgraph isomorphic matching to include edge labels in the mapping.
\item {\em match subgraph isomorph} uses an equality function on the definition of full subgraph isomorph to achieve exact matching of node and edge labels while maps node IDs using an injective function\footnote{Injective mapping of node IDs ensures structural match while equality of label mapping ensures that the graphs are identical even though the node IDs are different.}.
\end{itemize}
Among the above four modes of matching, structural subgraph
isomorphism is the least restrictive or selective, and so most
computationally expensive. While the match subgraph isomorphic
matching (in the literature it is known as labeled graph matching)
is the most restrcitive/selective, full and and label subgraph
isomorphism are increasingly less so. The idea here is that by
combining different selection and mapping constraints, called {\em
matching mode}, we can capture most popular graph matching concepts
and go beyond current definitions. Traditional deep equality
$\stackrel{d}{=}$ operator \cite{AbiteboulB95} in object-oriented
databases can be used to test if two graphs (or a subgraph) are
equal (or contained in the other graph) by requiring that node IDs
and labels be identical. Finally, by requiring that the two graphs
have equal number of nodes, we can also achieve graph isomorphism
for each case above.

\subsection{Main Motivation}

\begin{figure}[h]
\begin{center}
\resizebox{.49\textwidth}{1in}{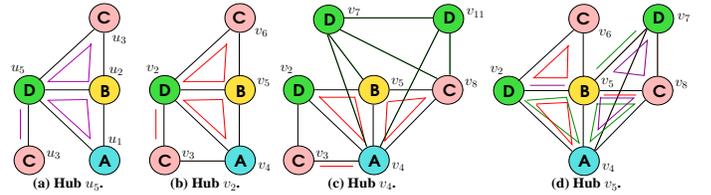}
\caption{Candidate generation for query graph hub $u_2$ on data
graph $d$.} \label{fig:cand}
\end{center}
\end{figure}

The vertex labeled\footnote{As will be discussed shortly, we can add
edge labels easily in our approach without breaking any definition
or requiring any change in our algorithms. Although the examples we
use to discuss the features of minimum hub cover and our graph
algorithms have undirected edges, a simple adjustment to the hub
representation can non-intrusively accommodate the directionality as
well.} graphs in figure \ref{fig:example} show two query graphs
$q_1$ and $q_2$, and a data graph $d$ respectively. In these graphs,
each node has a unique node ID such as $u_i$, $v_j$ and $w_k$, and a
label such as $A$, $B$ and $C$ (shown in uppercase with unique color
codes). If all the labels are empty, the graph is considered
unlabeled. If we matched query graph $q_1$ with the data graph $d$
for structural subgraph isomorphism, we will compute the green,
brown and red matches, among others, as we are only required to map
node IDs. However, if we were to compute match subgraph isomorphs,
only solution we can compute then is the green subgraph. If label
subgraph isomorphs were being sought instead, the solution will
include the brown subgraph where we map pink (C) to pink, green (D)
to blue (A), mustard (B) to mustard, and blue to green, along with
the obvious green subgraph. Let us illustrate this naive matching
process using figures \ref{fig:example} and \ref{fig:cand}.

Since there are six nodes in the graph $q_1$, a total of six hubs
are possible. Similarly, the graph $d$ can be represented by a total
of eleven hubs. Let us first match $q_1$ with $d$ in match subgraph
isomorphism mode. Under this constraint, for hub $u_5$ (shown
clock-wise rotated by $180^{\circ}{\rm C}$ in purple in figure
\ref{fig:cand}(a)), we can only find one hub, hub $v_2$ shown in
figure \ref{fig:cand}(b), in $d$ which is a supergraph of hub $u_5$
and can produce a structure identical to $u_5$ (shown as red edges)
on proper mapping of the node IDs. These structures are called
candidate graphlets. However, if we choose to match in structural or
label subgraph isomorphism mode, we can find more hubs as capable of
generating candidates structures. For example, in label subgraph
isomorphism mode, hubs $v_4$ and $v_5$ can also generate candidates
(possible candidate structures are shown in red, purple and green)
as shown in figures \ref{fig:cand}(c) and \ref{fig:cand}(d).

To complete the matching, we can continue to match all the other
hubs in $q_1$ in a similar way, and by applying the substitutions
generated for each set of previous matches to the candidate hubs to
eventually compute the green match as shown in figure
\ref{fig:example}(b). These graphlets can be joined or pieced
together in both bottom-up fashion using a process similar to
natural join, or in top-down manner using a depth-first
search\footnote{The complete top-down matching process is shown in
algorithm \ref{alg:find}. In this algorithm $\mu$ is the structure
mapping function that generates the mapping, $\theta$ is the
substitution list from previous steps, $[\theta]$ is the application
of the substitution, and $\bullet$ is the composition function of
two substitutions.}. In both cases, the graphlets that do not stitch
to form the target graph will eventually be eliminated.

Clearly, the dominant cost in this naive algorithm is in candidate
generation. Therefore, it would be prudent to seek opportunities to
curtail the candidates that do not have a realistic chance of
contributing to the result, or will produce redundant candidates.
For example, we can be smarter and choose to match $u_4$ or $u_5$
only without compromising the outcome. We can further speed up the
process by noticing that $u_4$ is a green D node and there are four
such nodes in $d$, all of which will generate a total of twelve
candidates even though only four will survive the node mapping, and
only one will join with the first candidate to complete the
computation. On the other hand, if we decided to use $u_3$, we will
generate six candidates ($v_3$ is not one of them) of which only one
will survive the mapping and eventually the response. Furthermore,
if we started initially with $u_4$, and then try to map $u_5$, the
number of candidates generated will be even higher although the
response computed will still be the same. Interestingly though, if
the query is $q_2$ (note the similarity of the two queries except
that $w_4$ is now purple), it is definitely better to start matching
with $w_4$, because there is only one candidate and it will fail to
produce the response in the next step, as expected.

These observations lead us to devise the following query processing
strategy. First, we reduce the number of hubs or graphlets in the
query graph that we must match based on a new notion of edge
covering in graph theory, called the {\em minimum hub cover} (MHC).
A minimum hub cover essentially means a subset of the nodes in a
graph accounts for all the edges in a graph. Secondly, the concept
of MHC helps exploit available meta-data on nodes to order the nodes
in priority order based on their selectivity to prune search space,
that we call a {\em query plan}. In ordering the nodes, we explore
the nodes that will most likely produce the least number of
candidates first\footnote{The ordering algorithm for query graphlets
is shown in algorithm \ref{alg:order}.}. Given the fact that a query
graph may have multiple MHCs, it also offers us the opportunity to
choose the best query plan for a database instance. Finally, we are
now able to use access structures such as hash index and set index
to find only the nodes that are relevant for expanding nodes at a
given point in a query plan. In fact, the matching algorithm
\ref{alg:find} uses two such indices $I_H$ and $I_S$. The query plan
can be implemented as a top-down or bottom-up procedure based on the
expected number of candidates and a choice can be made based on the
expected cost. In particular, it is also possible to reorder the
query plan in a top-down procedure to prune search space dynamically
in a way similar to best-first search.

\subsection{Organization of the Paper}

Covering based graph matching is proving to be an interesting and
emerging research direction although we are aware of only Sigma
\cite{MongioviNGPFS10s} which used set covering directly for
matching very small graphs, while \cite{ImamuraIT06,ChenK05}
indirectly used covering for graph matching tangentially. Our goal
in this paper, however, is to formally introduce the idea of graph
representation using graphlets and graph query processing using the
minimum hub cover of query graph graphlets\footnote{In our recent
research on a declarative graph query language called {\em NyQL}
\cite{NyQL-gdm-2012s}, we have informally introduced the idea of the
MHC and discussed how it can be exploited to represent graphs as
nested relations and develop graph query operators in a way similar
to the notion of the deep equality operator \cite{AbiteboulB95} in
object-oriented databases.}. Our focus is to convince the skeptics
that these two concepts help achieve the separation in graph
representation and storage, indexing, query plan generation, and
query optimization conveniently.

Once this model is accepted in principle, two main computational
problem emerge both of which are computationally hard -- computing
MHC and graph matching using subgraph isomorphism as the primary
vehicle. In this paper, we only address the first issue, that is the
computational aspects of MHC.  But for the sake of completeness, we
also briefly present an outline of the cost-based optimization
strategy for the ordering of graphlets in the MHC as a candidate
query plan, and a query processing algorithm that uses indices for
the execution of a query plan. By doing so, we demonstrate that
cost-based query optimization is feasible if we are able to compute
the MHC of a query graph. Finally, we believe that even if
algorithms such as SUMMA \cite{ZhangLY10}, NOVA \cite{NOVA}, TALE,
and SAPPER do not use a notion similar to hubs as we do, they do use
the notion of neighborhood and will benefit from the development
presented in this paper if they consider a similar covering, i.e.,
edge or vertex covers, of the queries. The results in this paper
then becomes directly relevant to those research as well because we
show how the cost of covering computation may vary depending on the
type and parameters of the graphs being considered.

The remainder of the paper is organized as follows. We discuss
background of the research related to MHC in section
\ref{sec:related}. In this section, we also discuss related research
in covering computation based on which we formulate our
characterization of MHC. The formal treatment of MHC and its
application in query plan generation is discussed in section
\ref{sec:prel}. Similar to other covering problems such as set
cover, and minimum vertex cover, MHC turns out to be an NP-complete
problem as well. Therefore, it can be framed as an optimization
problem and made a candidate for heuristic solutions. In section
\ref{sec:MHC}, we discuss an integer programming formulation of the
MHC problem as a prelude to our main results on its computability.
We have implemented the algorithm using the IBM ILOG optimization
engine CPLEX. The experimental results in section \ref{subsec:eval}
based on the design in section \ref{sec:comp} suggest that solving
MHC to optimality is not a concern for many graph types. A summary
of interesting and possible future research issues that are still
outstanding is discussed in section \ref{sec:fut}. We finally
conclude in section \ref{sec:conl}.

\section{Background and Related Research}
\label{sec:related}

In our earlier research on IsoSearch
\cite{RINQL-sac-2011-IsoSearchs}, we have shown that the notion of
{\em structural unification} helps to extract all possible matches
of two hubs under a mapping function or a substitution list. While
this atomic matching process generates a potentially large candidate
pool, we were able to avoid the large cost related to testing for
conformity of the candidate target structure with that of the query
graph that most other algorithms incur. In our case, conformity is
an eventuality and automatic if a match exists. We have also shown
that IsoSearch performs significantly better than traditional
algorithms such as Ullmann and VFLib, have a significantly low
memory footprint, and is able to handle arbitrary sized query and
data graphs (because we handle only pairs of graphlets at a time).
The concepts of hubs and minimum hub covers also help model various
definitions of exact graph matching along the lines of
\cite{TrisslL07,ZhangLGZ09,GADDI}, as well as approximate graph
matching in the spirit of TALE.

Tangentially to this research, in our recent top-$k$ graph matching
algorithm TraM \cite{TraM-TCBB-2012s}, we have explored the idea of
hub matching as a unit of comparison and computed structural
distance of attributed hubs without the need for explicit use of
indices. In this approach, we have developed a quantification for a
hub's structural feature as a random walk score \cite{BoldiSV07}.
Since random walk scores encompass the global topological properties
of a node as a hub, from the standpoint of graph matching, it can be
used to compare topological orientation and relative importance of
graph nodes. These scores thus effectively capture the topological
likeness and structural cues shared among the hubs, and were
effectively exploited for approximate graph matching in TraM. A
similar method was used in \cite{FoussPRS07s} to compute computation
of similarities of nodes of a graph for collaborative
recommendation. The random walk based approach, however, does not
offer much opportunity for cost-based query optimization based on
the evolving states of the database extension in ways similar to
\cite{ZhaoH10,TribleS2007} because it is largely similar in nature
to many algorithmic counterparts such as SUMMA, NOVA, TALE, and
SAPPER.


As we shall elaborate in the subsequent part of this section, the
MHC problem is closely related to two well-known combinatorial
optimization problems, the \textit{set covering problem} (SCP) and
the \textit{minimum vertex cover} (MVC) problem, which are, in turn,
share a similar mathematical programming model. We next discuss the
relationship between the MHC problem and the MVC problem, and then
tie this discussion to a general set covering formulation that we
also adopt in this work.

Let us start with the integer programming (IP) model for the MVC
problem:
\begin{align}
  \mbox{minimize } \ &\sum_{j\in V} x_{j}, & \label{VC obj function}\\
  \mbox{subject to } \ & x_i+x_j \geq{1} & (i,j) \in E, \label{VC coverage}\\
  &\ x_{j}\in\{0,1\},& j\in V, \label{VC integrality}
\end{align}
where $x_j$ is a binary variable that is equal to 1, if vertex $j$
is in the cover and 0, otherwise. The objective function \eqref{VC
obj function} evaluates the total number of vertices in the cover.
Constraints \eqref{VC coverage} ensure that every edge is covered by
at least one vertex, and constraints \eqref{VC integrality} enforce
binary restrictions on the variables. To give a concrete example,
suppose that we are trying to find the optimal MVC in the graph
shown in figure \ref{fig:example}(a). The corresponding IP model is
then given by

\begin{alignat}{7}
 \mbox{minimize} ~~   x_1 &{} + &{} x_2 &{} + &{} x_3 &{} + &{} x_4 &{} + &{} x_5 &{} + &{} x_6 &{},\nonumber\\
 \mbox{subject to} ~~ x_1 &{} + &{} x_2 &{}   &{}     &{}   &{}     &{}   &{}     &{}   &{}     &{} \geq{1}, \ \ \ [(u_1, u_2)]\nonumber\\
                      x_1 &{}   &{}     &{}   &{}     &{}   &{}     &{} + &{} x_5 &{}   &{}     &{} \geq{1}, \nonumber \ \ \ [(u_1, u_5)]\\
                          &{}   &{} x_2 &{} + &{} x_3 &{}   &{}     &{}   &{}     &{}   &{}     &{} \geq{1}, \nonumber \ \ \ [(u_2, u_3)]\\
                          &{}   &{} x_2 &{}   &{}     &{}   &{}     &{} + &{} x_5 &{}   &{}     &{} \geq{1}, \nonumber \ \ \ [(u_2, u_5)]\\
                          &{}   &{}     &{}   &{} x_3 &{} + &{} x_4 &{}   &{}     &{}   &{}     &{} \geq{1}, \nonumber \ \ \ [(u_3, u_4)]\\
                          &{}   &{}     &{}   &{} x_3 &{}   &{}     &{} + &{} x_5 &{}   &{}     &{} \geq{1}, \nonumber \ \ \ [(u_3, u_5)]\\
                          &{}   &{}     &{}   &{}     &{}   &{}     &{}   &{} x_5 &{} + &{} x_6 &{} \geq{1}, \nonumber \ \ \ [(u_5, u_6)]\\
                      x_1 &{} , &{} x_2 &{} , &{} x_3 &{} , &{} x_4 &{} , &{} x_5 &{} , &{} x_6 &{} \in\{0,1\}. \nonumber
\end{alignat}
The model minimizes the total number of selected vertices while
satisfying the coverage constraints written for each edge. For the
sake of clarity, the edge corresponding to a constraint is
designated at the end of each line within the brackets. The first
constraint, for example, implies that the edge $(u_1, u_2)$ can be
covered by vertices $u_1$ and $u_2$. Since an edge can be covered
only by the vertices incident to it, each constraint in the IP
formulation of the MVC problem involves exactly two variables. In
this particular example, $\{u_2,u_5\}$ is the unique optimal
solution.

When it comes to the mathematical programming model of the MHC
problem, we need to pay attention to the fact that a vertex (as a
hub) covers not only the edges incident to itself but also those
edges between its immediate neighbors. Using this fact, we obtain
the following IP formulation of the MHC problem for the graph in
figure \ref{fig:example}(a):
\begin{alignat}{7}
 \mbox{minimize} ~~   x_1 &{} + &{} x_2 &{} + &{} x_3 &{} + &{} x_4 &{} + &{} x_5 &{} + &{} x_6 &{},\nonumber\\
 \mbox{subject to} ~~ x_1 &{} + &{} x_2 &{}   &{}     &{}   &{}     &{} + &{} x_5 &{}   &{}     &{} \geq{1}, \nonumber\\
                          &{}   &{} x_2 &{} + &{} x_3 &{}   &{}     &{} + &{} x_5 &{}   &{}     &{} \geq{1}, \nonumber \\
                          &{}   &{}     &{}   &{} x_3 &{} + &{} x_4 &{}   &{}     &{}   &{}     &{} \geq{1}, \nonumber \\
                          &{}   &{}     &{}   &{}     &{}   &{}     &{}   &{} x_5 &{} + &{} x_6 &{} \geq{1}, \nonumber \\
                      x_1 &{} , &{} x_2 &{} , &{} x_3 &{} , &{} x_4 &{} , &{} x_5 &{} , &{} x_6 &{} \in\{0,1\}. \nonumber
\end{alignat}
Notice that unlike the IP formulation of the MVC problem, the number
of constraints reduces since multiple edges can be covered by the
same set of vertices. For instance, the second constraint shows that
vertices $u_2$, $u_3$ and $u_5$ cover the edges $(u_2,u_3),
(u_2,u_5)$, and $(u_3,u_5)$. As a consequence of this hub property,
the number of variables appearing in a constraint is greater than or
equal to two. In fact, this number can easily go up to the number of
vertices, because the vertices incident to an edge may be connected
to all other vertices forming an abundant number of triangles.
Clearly, the cardinality of the MHC can be far less than that of the
cardinality of the MVC due to the additional non-incident edges
covered by those vertices in a triangle. Therefore, for
triangle-free graphs, the optimal solutions for the MHC problem and
the MVC problem naturally coincide.

\subsection{Minimum Hub Cover: As a Special Case of Set Covering}

The above formulations of the MVC and MHC problems actually show
that both problems are just the special cases of the SCP. Given a
fixed number of items and a family of sets collectively including
(covering) all these items, the objective of the SCP is to select
the least number of sets (minimum cardinality collection) such that
each item is in at least one of these selected sets. If an edge
corresponds to an item and a set is formed with the edges that can
be covered by each vertex, then the connection between the SCP and
the MHC problem as well as the MVC problem can easily be
established. To formalize this discussion, below we give the generic
IP formulation for the MHC problem:
\begin{align}
  \mbox{minimize } \ &\sum_{j\in V} x_{j}, & \label{HC obj function}\\
  \mbox{subject to } \ & x_i+x_j+\sum_{\substack{(i,k)\in E\\(j,k)\in E}} x_k \geq{1}, & (i,j) \in E, \label{HC coverage}\\
  &\ x_{j}\in\{0,1\},& j\in V. \label{HC integrality}
\end{align}
Again, the binary variable $x_j$ is equal to 1, if vertex $j$ is in
the hub cover, and the objective is to minimize the number of
vertices used in the cover. Constraints \eqref{HC coverage} ensure
that every edge is covered by at least one hub node in the cover.
Finally, the constraints \eqref{HC integrality} enforce the binary
restrictions on the variables. Although the number of constraints
seems equal to the number of edges, we remind that multiple edges
can be covered by the same set of vertices (see MHC example above).

The introduction of the hub cover concept to the literature is quite
recent \cite{NyQL-gdm-2012s}. Thus, to the best of our knowledge,
the solution methods for the MHC problem have not been examined in
the literature. However, closely related problems, the MVC problem
and the SCP, have been extensively studied before. Take the MVC
problem; approximation algorithms \cite{Gomes06,Halperin02},
heuristic solutions \cite{Balaji10}, evolutionary algorithms
\cite{Khuri94,Evans98} can be listed among those numerous solution
methods. The SCP is no different. From approximation algorithms
\cite{Caprara00,Gomes06} that have good empirical performances to
randomized greedy algorithms \cite{Feo89,Haouari02}, and from local
search heuristics \cite{Lan07,Yagiura06} to different
meta-heuristics \cite{Caserta07,Ren10,Azimi10} have been proposed
for solving the SCP.

\section{Minimum Hub Cover: The Formal Model}
\label{sec:prel}

As mentioned earlier, in graph $q_1$, node or hub $u_5$ covers the
edges $(u_1,u_5)$, $(u_1,u_2)$, $(u_2,u_5)$, $(u_2,u_3)$,
$(u_3,u_5)$ and $(u_5,u_6)$. Similarly, the hub $u_3$ covers edges
$(u_3,u_2)$, $(u_5,u_2)$, $(u_3,u_5)$ and $(u_3,u_4)$. Since the set
of hubs $\{u_5,u_3\}$ covers all the edges of $q_1$, this set is
called a {\em hub cover} of $q_1$, denoted $cover(q_1)$, and so are
the sets $\{u_5, u_4\}$, $\{u_2,u_6,u_4\}$, and $\{u_1,u_5, u_4\}$.
However, for the purposes of graph matching, it is sufficient if we
matched the set $\{u_5,u_3\}$ with another graph as any super set
this will not add any more new node or edge matching. We thus call
this set the {\em minimum hub cover} of $q_1$, denoted $MHC(q_1)$
and formalized in the following definition.

\begin{definition}[Minimum Hub Cover]
{\em For a given graph $G=<V, E>$, $M\subseteq V$ is a minimum hub
cover of $G$, denoted $MHC(G)$, if $M$ is the smallest set, and for
every edge $<s,d>\in E$, either $d\in M$ or $s\in M$, or there
exists edges $<s,c>\in E$ and $<d,c>\in E$, and $c\in M$. The set of
all $MHC(G)$ is denoted as $\Gamma(G)$.}
\end{definition}

It should be apparent that given a graph $G=\langle V, E\rangle$,
the number of hub nodes in $M$ lies within the range $1 \leq |M|\leq
\frac{|V|}{2}$. For example, both $\{u_5,u_3\}$ and $\{u_5,u_4\}$
are in $\Gamma(q_1)$, and $|\{u_5,u_3\}|= |\{u_5,u_4\}|$. It should
also be apparent that hubs in a given MHC ensures edge connectivity
because all edges are covered. But it is likely that a given
sequence of hubs may not retain edge connectivity, and thus ordering
of the elements in $MHC(G)$ matters and as such this ordering
relationship is a key to processing queries efficiently.
Intuitively, the query optimization approach we present essentially
is the problem of computing the set $\Gamma(Q)$ when computable,
then ordering the elements in each set $c_i\in \Gamma(G)$ using a
cost function $\chi$ (discussed in detail in section \ref{sec:cost})
such that $\chi(c_i)$ is minimized, and then choose MHC $c_j$ such
that $\chi(c_j)$ is the minimum among all MHCs in $\Gamma(Q)$ under
$\chi$. The least cost ordering relationship implied by $\chi(c_j)$
is then taken as the query plan to be executed using a depth-first
matching algorithm.

\subsection{Hubs as Graph Representation}

Traditionally, graphs are represented as a pair $\langle V,
E\rangle$ where $V$ is a set of vertices and $E$ is a set of edges
over $V$. Such a representation does not carry any structural
information of which vertices are a part of, and they are not
visible until structures are constructed from the set of edges. To
ease computational hurdles and aid analysis, some models have used
vertices and their neighbors as a unit of representation
\cite{tales}, i.e., $r_v=\langle v, N\rangle$ where $v$ is vertex in
$V$, and $N$ is a set of neighbors such that $(v,n)\in E \Rightarrow
n\in N$. A graph is then modeled as a set of such units. While this
representation captures some structural cues, it still is pretty
basic.

The decision on the degree of structural information that can be
captured and exploited is very hard. In one extreme, we have the
traditional representation $\langle V, E\rangle$ with no structural
information other than the edges, and the GraphQL model where pretty
much the entire graph structure is represented as a single XML
document. While GraphQL offers most selectivity due to its
representation, taking advantage of this in cost-based query
optimization is extremely difficult because meta-data now has to be
at the largest structural level. The models that are in between are
the star structures \cite{tales} and dynamically mined frequent
feature structures \cite{YanYH05s,ZouCZLL08} which we believe can
generally be difficult to index and maintained with the evolution of
the database.

We believe representing a graph as a set of hubs is a prudent
compromise because it assures a deterministic model, and yet offers
a realistic chance of efficient storage and processing of graph
queries. It is deterministic because each hub can be represented as
a triple of the form $r_v=\langle v, N_v, B_v\rangle$, called a {\em
graphlet}, where $v\in V$ is a node in graph $G=\langle V,
E\rangle$, and $N_v\subseteq V$, and $B_v\subseteq E$ such that for
each $v\in V$ all $n\in N_v$ are its immediate neighbors, and every
edge $b\in B_v$ are edges involving neighbors in $N_v$\footnote{We
follow this convention of representing a graphlet, equivalently
called a node or hub, throughout the paper unless stated
otherwise.}. For unlabeled and undirected graphs, this
representation model is sufficient. But for labeled and directed
graphs, this simple model can be extended without any structural
overhaul\footnote{For example, a hub of a node labeled undirected
graph can be represented simply as $r_v=\langle v, L_v, N_v,
B_v\rangle$ where $L_v$ additionally represents the node label. The
hubs in a fully labeled graphs can be modeled as yet another
extension as $r_v=\langle v, L_v, N_{v_l}, B_{v_l}\rangle$ where
$N_{v_l}$ are a set of pairs $(n,n_l)$ and $B_{v_l}$ are triples
$(n_1,n_2,e_l)$ such that $n_l$ and $e_l$ are edge labels for edges
between the hub and the neighbors, and among the neighbors
respectively. The directionality of the edges can be captured by
partitioning the sets $N_v$ and $B_v$ to imply directions. For
example, the expression $\langle v, (N_{v_t},N_{v_f}),
(B_{v_t},B_{v_f})\rangle$ means that (i) there are edges from $v$ to
every node in $N_{v_t}$, and from $N_{v_f}$ to $v$, and (ii) for
each edge $(n_1,n_2)$ in $B_{v_t}$, the sink node is $n_2$, and for
edges in $B_{v_f}$, it is reversed.}. We can now simply define a
graph as a set of graphlets, i.e., $G = \bigcup_{\forall v, v\in V}
r_v$. In fact, in NyQL, we have shown that such a representation can
be conveniently captured using traditional storage structures such
as nested relations or XML documents.

\subsection{Cost Function $\chi$ and Graphlet Ordering Relation $\prec$}
\label{sec:cost}

Given two graphlets $r_v$ and $r_u$, and a graph $D$, we define an
ordering relation $r_v \preceq r_u$, called the {\em selectivity
ordering}, between them to mean $r_v$ precedes $r_u$ (or $r_v$ costs
less than $r_u$ in processing time) in two principal ways.
\begin{itemize}
\item First, when the selectivity of $r_v$, denoted $sel(r_v,D)$, is
  higher than $sel(r_u,D)$. Technically, $sel(x,G)$ means the number of graphlet
  instances $r_y$ in graph $G$ that satisfy the condition $|N_y|\geq
  |N_x|$ and $|B_y|\geq |B_x|$, and selectivity of $r_v$ is higher
  than that of $r_u$ in a graph $G$ if $sel(r_v,D) \leq
  sel(r_u,D)$. We call it {\em node selectivity}, denoted $\preceq_v$.
\item Second, when the selectivity of $N_v$ and $B_v$, called the {\em structural selectivity} and denoted $\preceq_s$, is higher than the selectivity of $N_u$ and $B_u$, denoted respectively as $sel(N_v,B_v, D)$ and $sel(N_u,B_u, D)$, i.e., $sel(N_v,B_v, D)\leq sel(N_u,B_u, D)$, where $sel(N_x,B_x, D)$ is the number of graphlet instances $r_y$ in graph $G$ that satisfy the condition $|N_y|\geq |N_x|$ and $|B_y|\geq |B_x|$.
\item Finally, the {\em neighbor selectivity} (denoted $\prec_n$) $sel(N_v, D)\leq sel(N_u, D)$ holds where $sel(N_x, D)$ represents the number of graphlet instances $r_y$ in graph $G$ that satisfy the condition $|N_y|\geq |N_x|$.
\end{itemize}

\begin{figure}[h]
\centering \subfigure[Expansion of top
node.]{\label{top}\resizebox{.45\textwidth}{.75in}{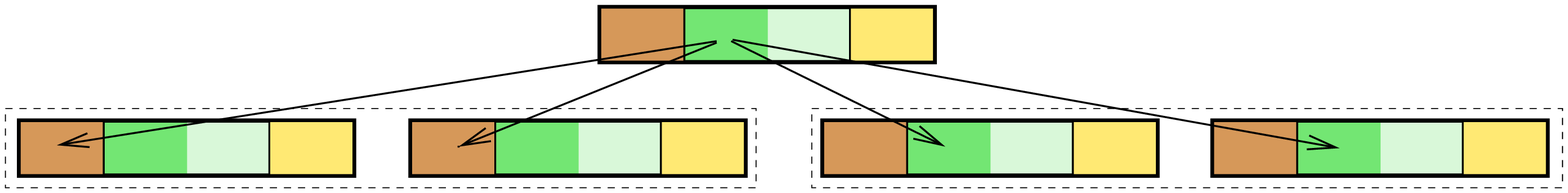}}
\subfigure[Dot and star expansion of
$Q$.]{\label{dot-star}\resizebox{.45\textwidth}{.75in}{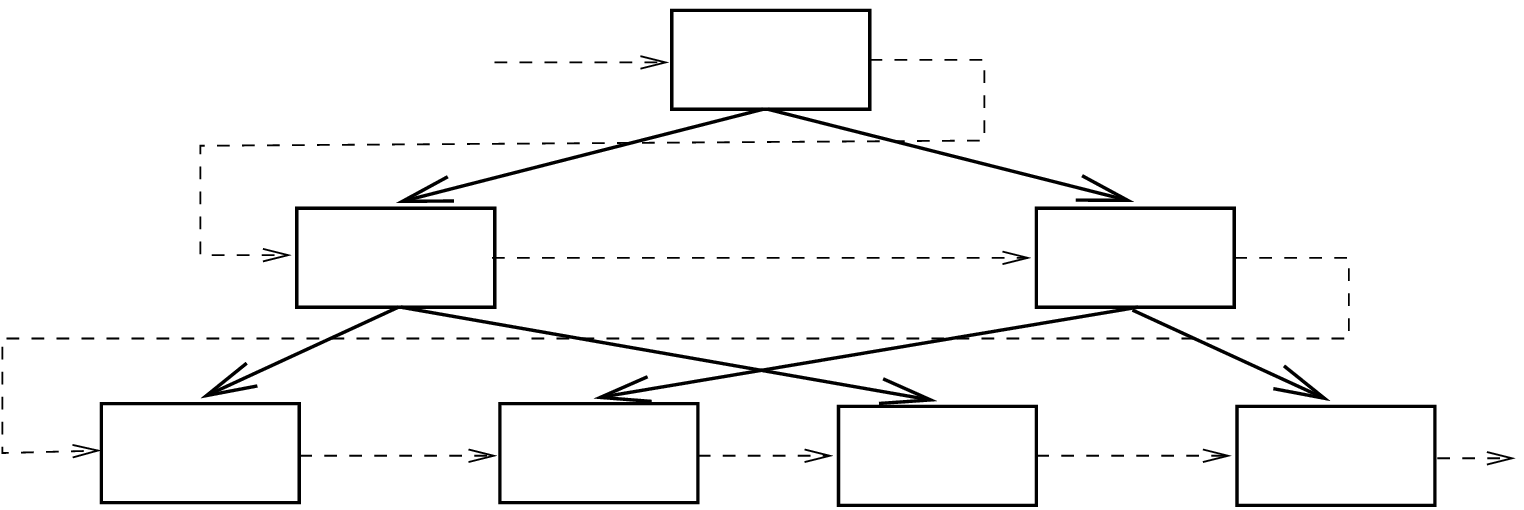}}
\caption{Ordering $Q$ based on expected cost using selectivity.}
\label{g-exmp2}
\end{figure}

The query plan as the total ordering $\prec$ then can be induced
from the precedence relationship $\preceq$ of graphlets in $G$ using
the procedure {\bf Graphlet Ordering} in algorithm \ref{alg:order}.
Algorithm \ref{alg:order} is intuitively explained in figure
\ref{g-exmp2} where we show that the dot nodes are identified first
followed by the star nodes and ordered in terms of their
selectivity. In figure \ref{top}, the three components of a graphlet
are shown in brown, green and yellow, where the neighbors $N_v$ are
partitioned in to two sets $mapped(N_v)\subseteq N_v[\theta]$ and
unmapped subsets (shown in uppercase). Since an instantiated dot
node may only fetch a single hub at the most, this ordering is
likely to result in a least cost plan.

\IncMargin{.15em}
\begin{algorithm}[ht]
\KwIn{A graph $G$, a cost estimator $\chi$, a matching mode and data
dictionary.} \KwOut{A total ordering $\prec$ on $G$.}
Set $i=1$.\\
Choose the most selective graphlet as $r_{v_i}$ using node selectivity.\\
\While{not all graphlet in $\prec$ marked dead}
    {
    \ForEach{unmarked graphlet $r_v$ in $\prec$ order}
    {\While{$\exists r_x, r_{x}\in G$ not marked dead and $x\in N_{v}$ (i.e., $x$ is an immediate neighbor of $v$)}
        {\For{$\forall r_y, r_y\in N_{v}$}
            {{\If{$r_x\preceq_v r_y$ or $r_y\preceq_v r_x \wedge r_x\preceq_s r_y$ holds}
                {Increment $i$.\\
                Let $r_{v_{i+1}}=r_x$.\\
                Mark $r_{v_{i+1}}$ as a dot node, i.e., $r_{v_{i+1}}^\bullet$.}
            }
        }
    }
    }
    \ForEach{unmarked graphlet $r_v$ in $\prec$ order}
    {\While{$\exists r_x, r_{x}\in G$ not marked dead and $N_{v}\cap N_{x}\not = \emptyset$ (i.e., $x$ is a 2-hop neighbor of $v$ and share the nodes $N_{v}\cap N_{x}$ as neighbors)}
        {\For{$\forall r_y, N_y\cap N_{v}\not = \emptyset$}
            {{\If{$r_x\preceq_n r_y$ or $r_y\preceq_n r_x \wedge r_x\preceq_s r_y$ holds}
                {Increment $i$.\\
                Let $r_{v_{i+1}}=r_x$.\\
                Mark $r_{v_{i+1}}$ as a star node, i.e., $r_{v_{i+1}}^\star$.}
            }
        }
    }
    Mark $r_v$ dead.
    }
} \Return{Order $\prec$.} \caption{Graphlet Ordering}
\label{alg:order}
\end{algorithm}
\DecMargin{.15em}

\subsection{Indices for Efficient Access}
\label{sec:index}

Let us also assume that we have a hash index $I_H$ that given a node
ID (i.e., $v\in V$), returns the graphlet corresponding to $v$,
i.e., $I_H(v)=r_v$. Furthermore, we also have another index $I_S$
similar to \cite{TerrovitisBVSM11} such that it returns a set of
graphlets $R$ given a set of nodes $V_q$, and an integer $n_c$,
i.e., $I_S(V_q,n_c)\subseteq G$\footnote{Recall that $G$ is now a
set of graphlets, or basically a set of nested tuples.}, such that
for each $r_v\in R$, $V_q\subseteq N$, and $|N|\geq n_c$. In other
words, $I_S$ returns all graphlets that contain the set of nodes in
$V_q$ and have a minimum of $|V_q|\leq n_c$ neighbors.

\subsection{Outline of a Query Processor}
\label{sec:qp-alg}

The graph query processor for exact matching is simple. It is
composed of a depth-first search algorithm {\bf Find Solutions} that
maps a sequence of graphlets in a query graph presented in $\prec$
order to a set of graphlets of a data graph using a composable term
mapping function $\mu$. Algorithm \ref{alg:find} is assumed to have
access to a data graph $D$, one of four graph matching modes, and
the indices $I_H$ and $I_S$.

\IncMargin{.15em}
\begin{algorithm}[ht]
\SetKwFunction{FindSolutions}{Find Solutions}
\KwIn{A stack $T$ with query graph $Q$ in $\prec$ order (initially
$r_{v_1}$ as the top element), and a substitution list $\theta$
(initially empty).} \KwOut{Set $sol$ of all matched subgraphs of
$D$.} \If{stack $T$ empty}
    {Include $Q[\theta]$ in $sol$.\\
    \Return.
    }
Let $r_v=pop(T)$.\\
Apply substitution $\theta$ to $r_v$, i.e., $r_v\leftarrow r_v[\theta]$.\\
\eIf{$r_v$ is not a dot node and a star node (it is the top node in
$\prec$ order)}
    {Retrieve all $r_x$ from $D$ such that $|N_x|\geq |N_v$ and $|B_x|\geq |B_v$.\\
    }
    {
    \eIf{$r_v$ is a dot node}
        {Retrieve $r_x\leftarrow I_H(v)$ from $D$.\\
        }
        {
        \If{$r_v$ is a star node}
        {Let $M=mapped(N_v)$.\\
        Retrieve all $r_x\in I_S(M,|N_v|)$
        }
    }
} \ForEach{$r_x$} {\eIf{exists one substructure of $r_x$ similar to
$r_v$}
    {
        \ForEach{substructure $r_y$ of $r_x[\theta]$ similar to $r_v$}
            {
                \FindSolutions{$T,\mu(r_y,r_v)\bullet\theta$}.
            }
    }
    {
        \Return.
    }
} \Return{$sol$} \caption{Find Solutions} \label{alg:find}
\end{algorithm}
\DecMargin{.15em}

\subsection{Computational Characterization of Minimum Hub Cover}
\label{sec:MHC}

We formally present the computational characterization of MHC in
terms of its complexity class, and using the decision version of the
MHC problem (MHC-D) show that the MHC problem is indeed difficult
and that it is NP-complete.

\begin{theorem}
MHC-D is NP-complete.
\end{theorem}

\begin{proof}
  To show that MHC-D is NP-Complete, we answer the following question:
  given a graph $G(V,E)$ and an integer $k \leq |V|$, does there exist
  a subset $C\subseteq V$ with $|V| \leq k$ such that for every edge
  $(s,d)\in E$, $s\in C$ or $d\in C$ or there exists a vertex $c\in C$
  such that both $(s,c)\in E$ and $(d,c)\in E$? First, we argue that
  MHC-D is in NP because for a given yes-instance and $C' \subseteq V$
  with $|C'| \leq k$, we can verify in polynomial time that every edge
  in $E$ is covered by $C'$. We now complete the proof by a reduction
  from the MVC problem on triangle free graphs \cite{Poljak74}. In
  triangle-free graphs, for all $(s,d)\in E$ there does not exist a
  vertex $c$ such that $(s,c)\in E$ and $(d,c)\in E$. Thus, either
  $s$, or $d$ must be in $C'$ to cover the edge $(s,d)$. Consequently,
  the MHC and the MVC of a graph are equivalent in this class of
  graphs and MVC is NP-complete for this class of graphs
  \cite{Poljak74}.
\end{proof}

\section{Experimental Design}
\label{sec:comp}

The NP-completeness of MHC speaks to the hardness of the solvability
in general. The optimization strategy and query plan generation
presented in section \ref{sec:prel}, however, calls for the
computation of all MHCs of a given query $Q$, i.e., $\Gamma(Q)$,
which is indeed a hard problem. In this section, our goal is to
design a set of experiments using different graph types, size and
graph density parameters to study how the solvability and the
quality of MHC solutions depend on these parameters. The goal is to
experimentally identify problem classes for which available
solutions are practical and acceptable, and the classes for which
new heuristic solutions are warranted. We therefore employ different
algorithms to show the trade-offs between optimality and computation
time over different graph types. Although our experiments and
analysis involve producing only one optimal solution for a graph,
the insights gained can be leveraged to develop new algorithms to
compute $\Gamma(Q)$ or to choose other strategies when computing
$\Gamma(Q)$ is infeasible.

The optimal linear programming (LP) and IP solutions are obtained by
{\tt ILOG IBM CPLEX 12.4} on a personal computer with an Intel Core
2 Dual processor and 3.25 GB of RAM. In all problem instances, the
upper limit on the computation is set at 3,600 seconds. The batch
processing of the instances is carried out through simple C++
scripts. Our data set includes a total of 830 instances. We have 5
different instances for each combination of a graph type, size, and
density parameter to be able to draw conclusions.

\subsection{Selected Graph Types and Problem Classes}
\label{subsec:probclass}

We have chosen to use the benchmark database graph instances in
\cite{Santo03} and our own synthetically generated data set for our
numerical study. This is a very large database of different graph
types and sizes designed specifically to test the sophistication of
(sub)graph isomorphism algorithms. Since we are using subgraph
isomorphism as a basic vehicle for graph matching, the instances
selected are thus representative of the class of queries we are
likely to handle when we solve the MHC problem. The descriptions of
the graph instances we have chosen from this collection are listed
below.

\paragraph{Randomly connected graphs} These graphs have no special
structure and the number of vertices range from 20 to 1000 ($|V|$=
20, 60, 100, 200, 600, 1000). The parameter $\eta$ denotes the
probability of having an edge between any pair of vertices.  Thus,
this parameter, in a sense, specifies the sparsity of a graph. In
the database, three different values of $\eta$ (0.01, 0.05, and
0.10) are considered. Our data set includes a set of graphs of
different sizes for each value of $\eta$.

\paragraph{Bounded valence graphs} The vertices of the graphs in this
class have the same degree (\textit{fixed valence}). The sizes of
the instances are similar to those of the problem class (a). We use
three different values of valence -- 3, 6 and 9 to obtain graphs of
different size and valence.

\paragraph{Irregular bounded valence graphs} These graphs are
generated by introducing irregularities in the problem class (b).
Irregularity comes from randomly deleting edges and adding them
elsewhere in the graph. With this modification, the average degree
is again bounded but some of the vertices may have higher degrees.
The sizes of the instances are similar to those of problem classes
(a) and (b).

\paragraph{Regular meshes with 2D, 3D, and 4D} In graphs with 2D, 3D,
and 4D meshes, each vertex has connections with 4, 6, and 8
neighbors, and the numbers of vertices range from 16 to 1024, 27 to
1000, and 16 to 1296, respectively. Similar to the problem classes
(a), (b), and (c), we have a set of graphs for each combination of
size and dimension.

\paragraph{Irregular meshes} As in class (c), irregular meshes are
generated by introducing small irregularities to the regular meshes.
Irregularity comes from the addition of a certain number of edges to
the graph. The number of edges added to the graph is $\rho\times
|V|$, where $\rho \in \{0.2, 0.4, 0.6\}$. The number of vertices is
exactly the same as in problem class (d).

\paragraph{Scale-free graphs} This problem class includes the graphs
that follow a power-law distribution of the form
\[
P(k)\sim ck^{-\alpha},
\]
where $P(k)$ is the probability that a randomly selected vertex has
exactly $k$ edges, $c$ is the normalization constant, and $2\leq
\alpha \leq 3$ is a fixed parameter. We employed the scale-free
graph generator of C++ Boost Graph Library. The generator (Power Law
Out Degree algorithm) takes three inputs. These are the number or
vertices, $\alpha$ and $\beta$. Increasing the value of $\beta$
increases the average degree of vertices. On the other hand,
increasing the value of $\alpha$ decreases the probability of
observing vertices with high degrees. The sizes of the instances
range from 20 to 1000; $|V| \in \{20, 60, 100, 200, 600, 1000\}$ to
be precise. We considered two values for $\alpha \in \{1.5, 2.5\}$
and three values of $\beta \in \{100\times |V|, 200\times |V|,
500\times |V|\}$. Graphs in social networks, protein-protein
interaction networks, and computer networks are examples of this
class.

\subsection{Solution Methods Used} \label{subsec:methods}

We choose three solution methods to compute MHC -- (i) an exact
method to solve the problem to optimality, (ii) adapt two
approximation algorithms from the vertex cover literature capable of
computing feasible solutions fast,  and (iii) a mathematical
programming-based heuristic originally proposed for solving the SCP.

\subsubsection{Exact algorithm} The IP formulation \eqref{HC obj
  function}-\eqref{HC integrality} is solved by an off-the-shelf
solver to optimality.  Since the MHC problem is shown to be NP-Hard,
this approach may have practical value only for
small-to-medium-scale graphs. However, it sets a definitive
benchmark for comparing the performances of various heuristics.

\subsubsection{Approximation and greedy algorithms} We implemented two
different approximation algorithms. First algorithm selects the
vertex with the highest degree at each iteration. The aim is to
cover as many edges as possible. Next, all covered edges as well as
the vertices in the cover are removed from the graph. The algorithm
ends when there is no uncovered edge in the graph. The algorithm is
called the \textit{$H(\Delta)$-approximation algorithm} (GR1) for
the MVC problem. Here, $\Delta$ is the maximum degree in the graph,
and $H(\Delta)$ is evaluated by
\[
H(\Delta)=1+1/2+\ldots+1/\Delta.
\]
The second algorithm (GR2), the \textit{2-approximation algorithm},
is an adaptation of \cite{Bar81} originally proposed for computing a
near-optimal solution for the MVC problem. Unlike the previous
algorithm, it selects an edge arbitrarily, then both vertices
incident to that edge are added to the cover.

\subsubsection{Mathematical programming-based heuristics}
Yelbay et al. \cite{Yelbay12} propose a heuristic (MBH) that uses
the dual information obtained from the LP relaxation of the IP model
of SCP.  They show the efficacy of the heuristic on a large set of
SCP instances. In their work, the dual information is used to
identify the most promising columns and then form a restricted
problem with those columns. Then, an integer feasible solution is
found by one of the two approaches. In the first approach (MBH), the
exact IP optimal solution is obtained by solving \textit{the
restricted problem}. In the second approach, a METARAPS \cite{Lan07}
local search heuristic (LSLP) is applied over those promising
columns. We use both of these approaches.

\section{Analysis of Experimental Results} \label{subsec:eval}

We focus on analyzing and understanding the MHC solution methods in
section \ref{subsec:methods} on the instances in section
\ref{subsec:probclass} in three different axes: (i) optimal
solvability of MHC, (ii) quality of the solutions, and (iii)
computational cost of optimal solution. These analyses are aimed at
understanding which problem classes are inherently more difficult
relative to others so that depending on the application and query, a
suitable algorithm can be selected to compute MHC. We also discuss
the factors that increase the complexity of the problems.

\subsection{Optimal Solvability of Minimum Hub Covers}

\begin{figure}[t!]
\begin{center}
\subfigure[Randomly connected graphs]{\label{fig:randopt}
\includegraphics [height=1.23in, width=.23\textwidth]{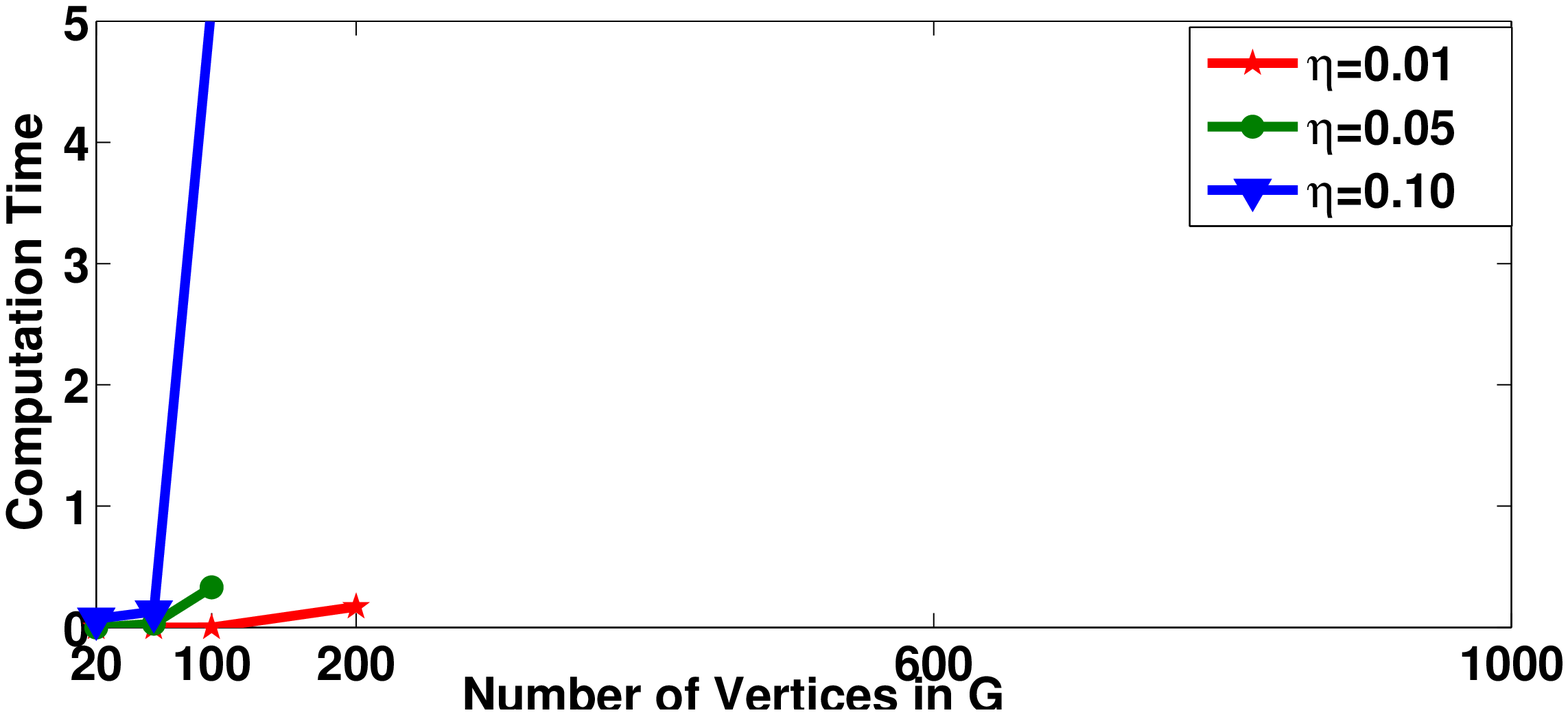}}
\subfigure[Bounded valence graphs]{\label{fig:boundopt}
\includegraphics [height=1.25in, width=.23\textwidth]{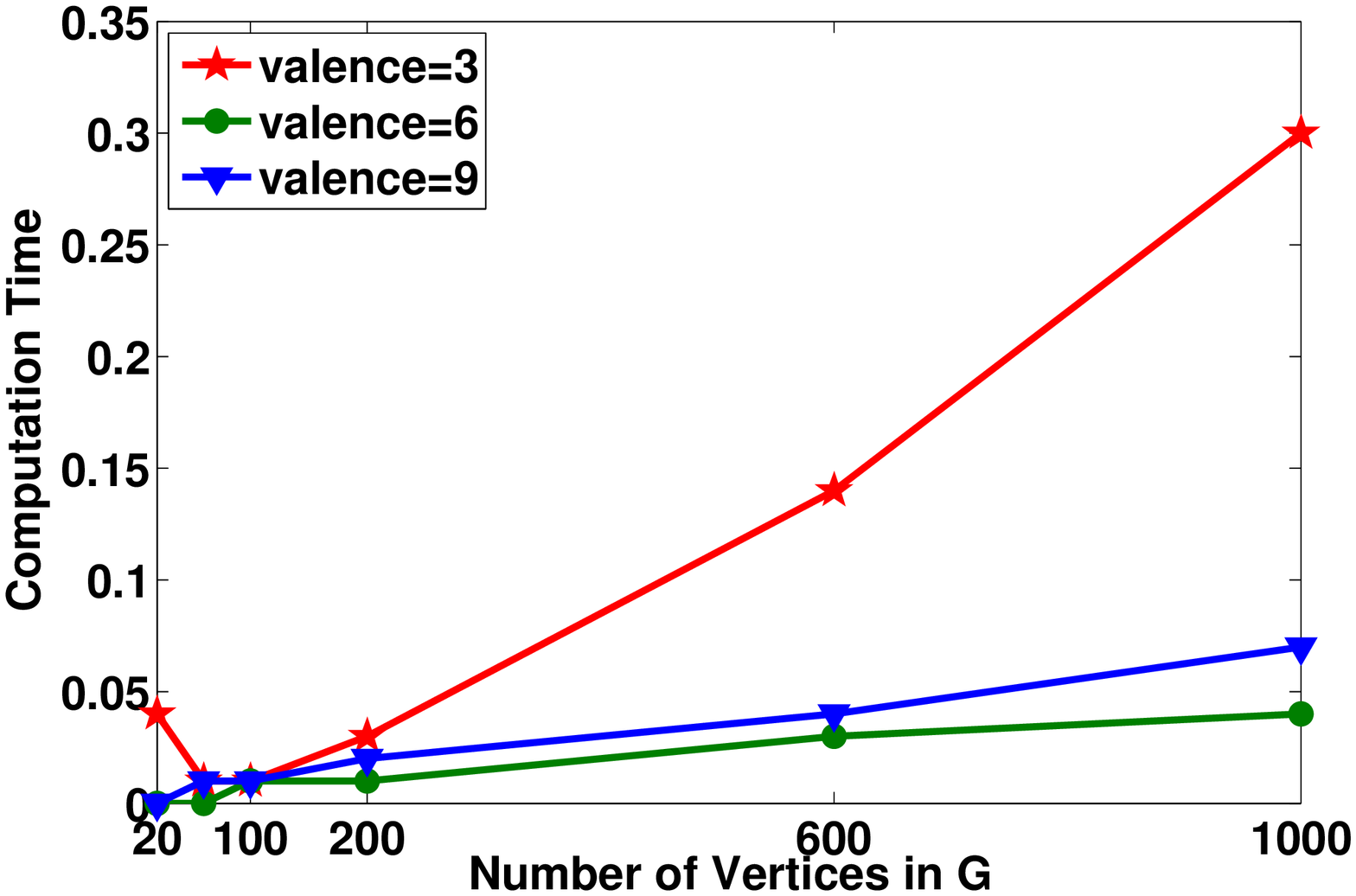}}
\subfigure[Irregular bounded valence graphs]{\label{fig:irboundopt}
\includegraphics [height=1.25in, width=.23\textwidth]{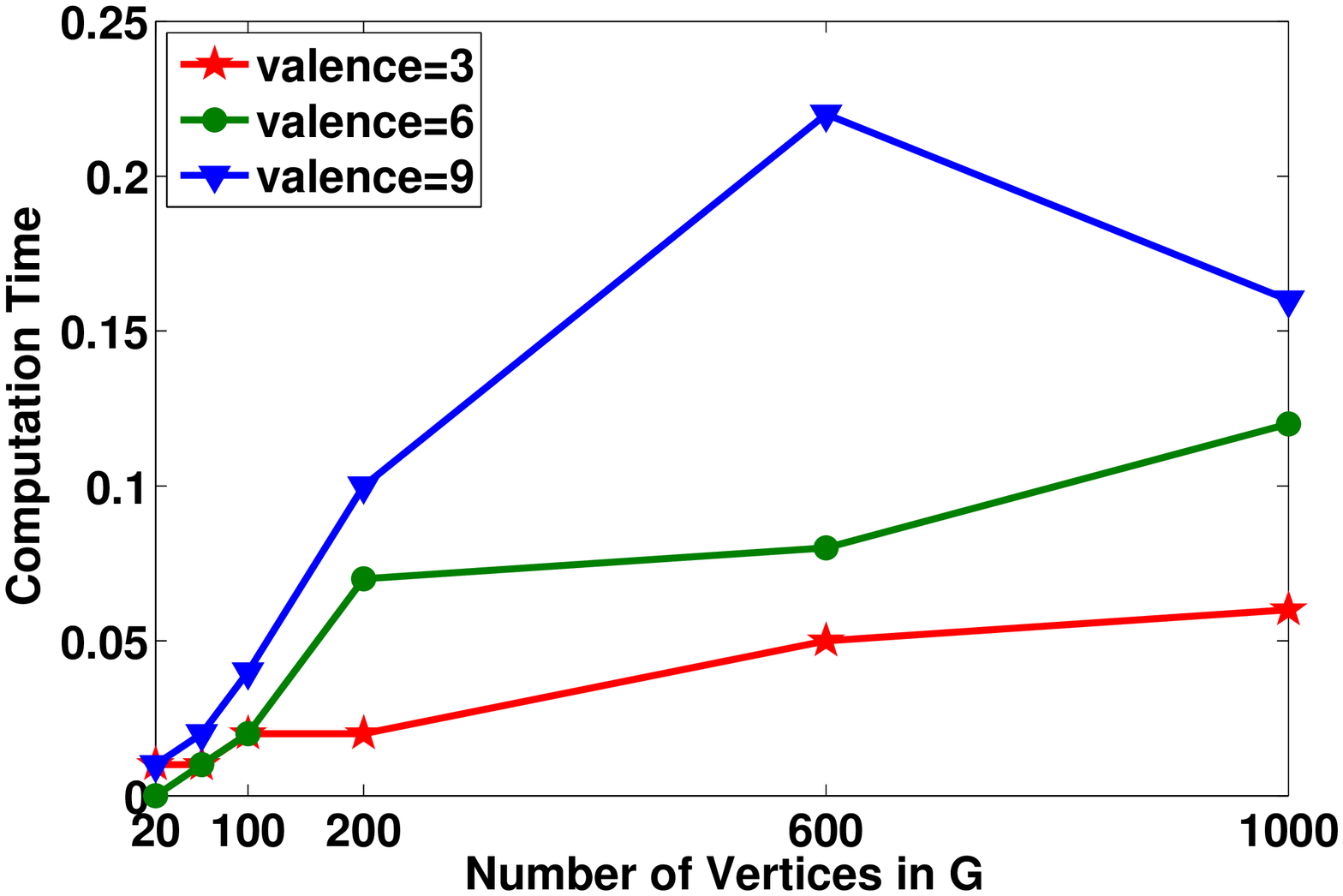}}
\subfigure[Regular meshes with 2D, 3D and 4D]{\label{fig:meshopt}
\includegraphics [height=1.25in, width=.23\textwidth]{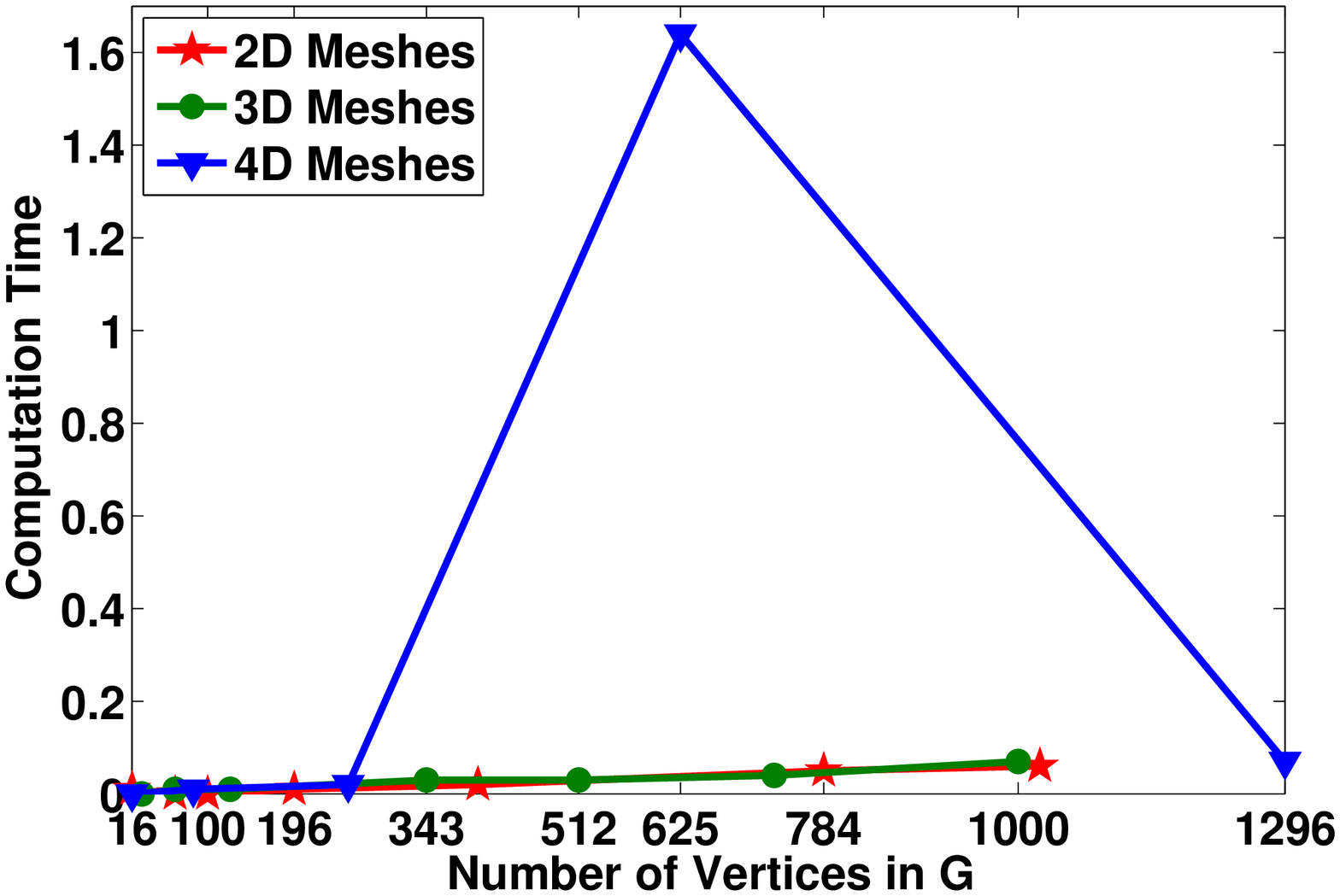}}
\subfigure[Irregular meshes with 2D]{\label{fig:irmeshopt2}
\includegraphics [height=1.25in, width=.23\textwidth]{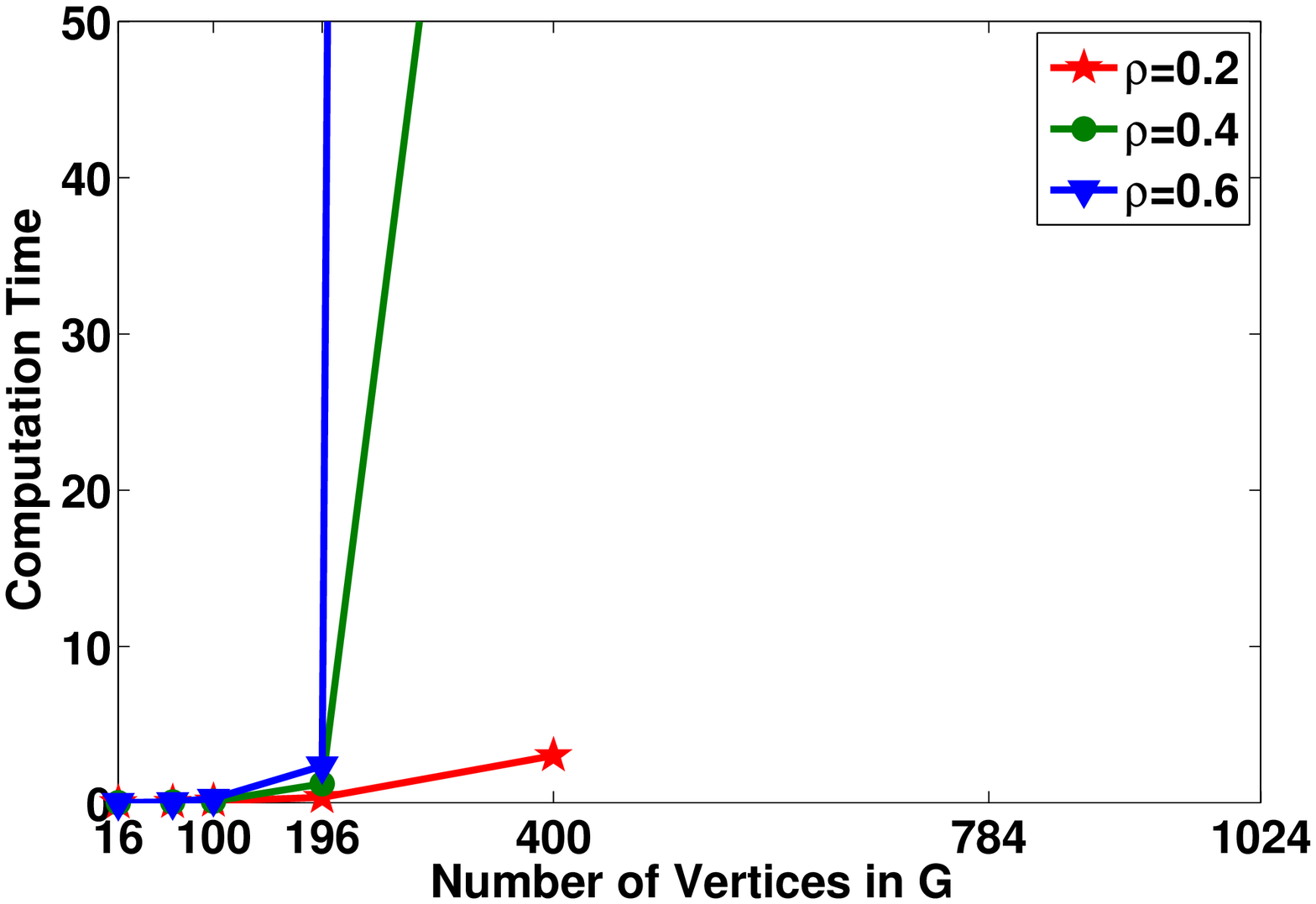}}
\subfigure[Irregular meshes with 3D]{\label{fig:irmeshopt3}
\includegraphics [height=1.25in, width=.23\textwidth]{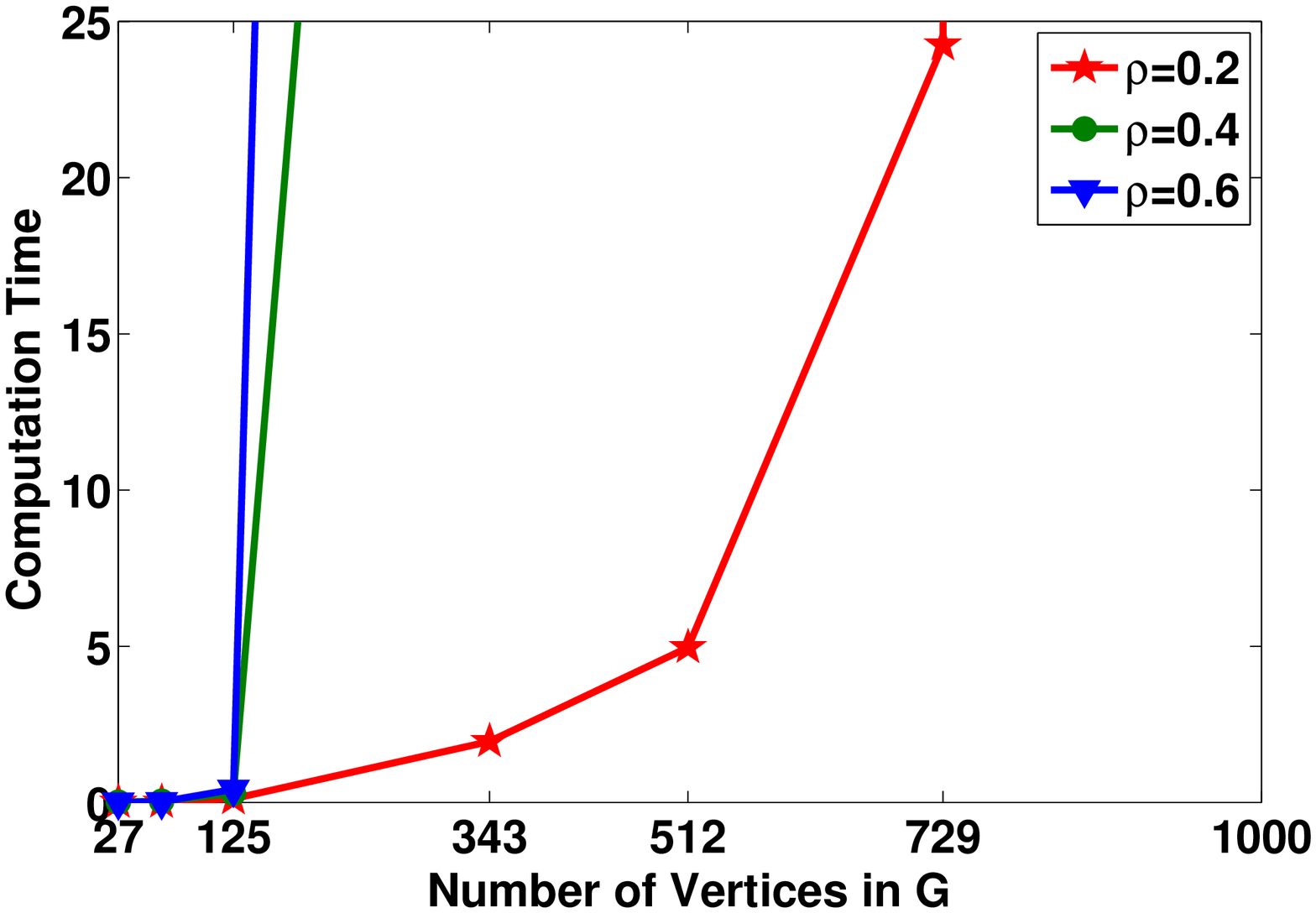}}
\subfigure[Irregular meshes with 4D]{\label{fig:irmeshopt4}
\includegraphics [height=1.25in, width=.23\textwidth]{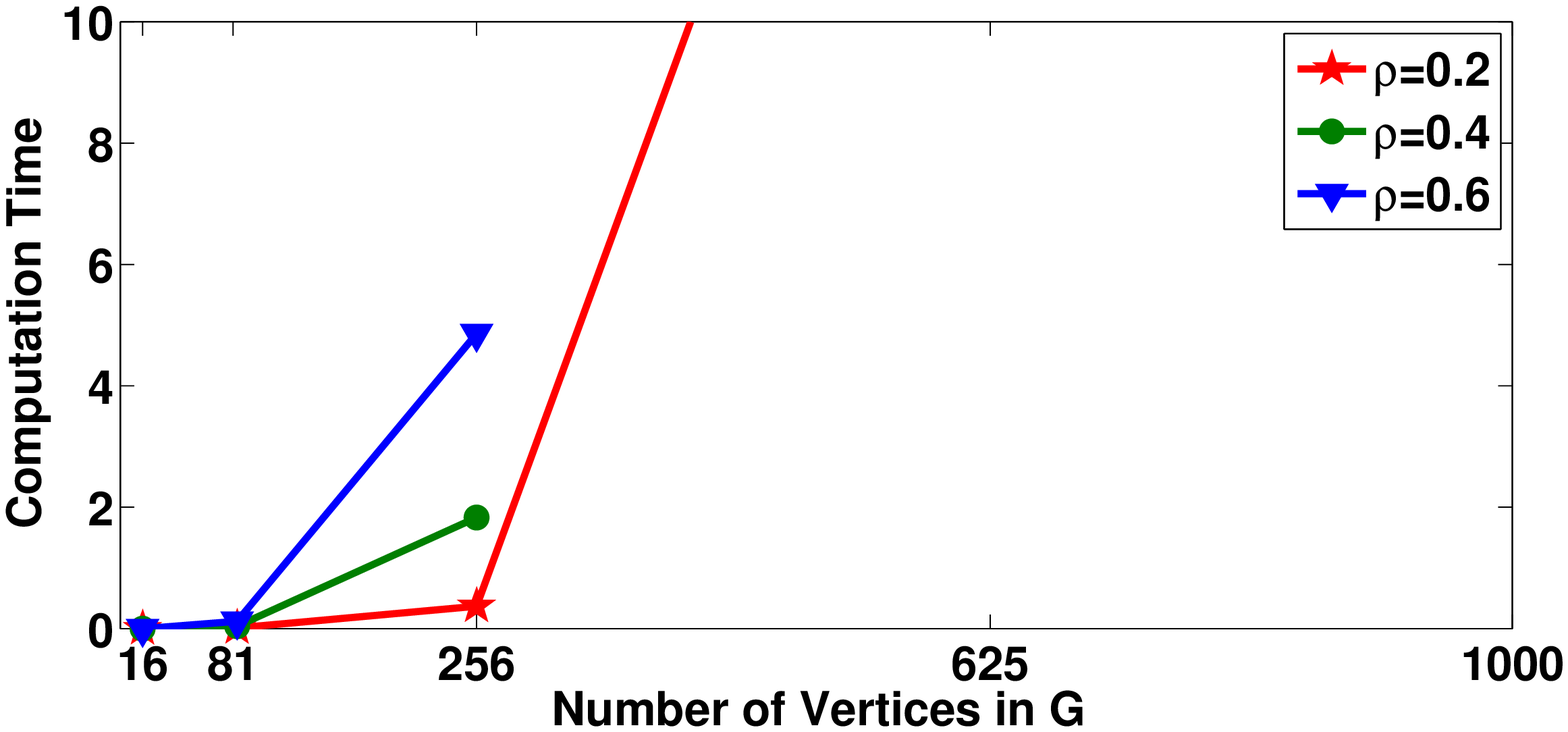}}
\subfigure[Scale-free graphs with
$\alpha=1.5$]{\label{fig:scalefree1}
\includegraphics [height=1.25in, width=.23\textwidth]{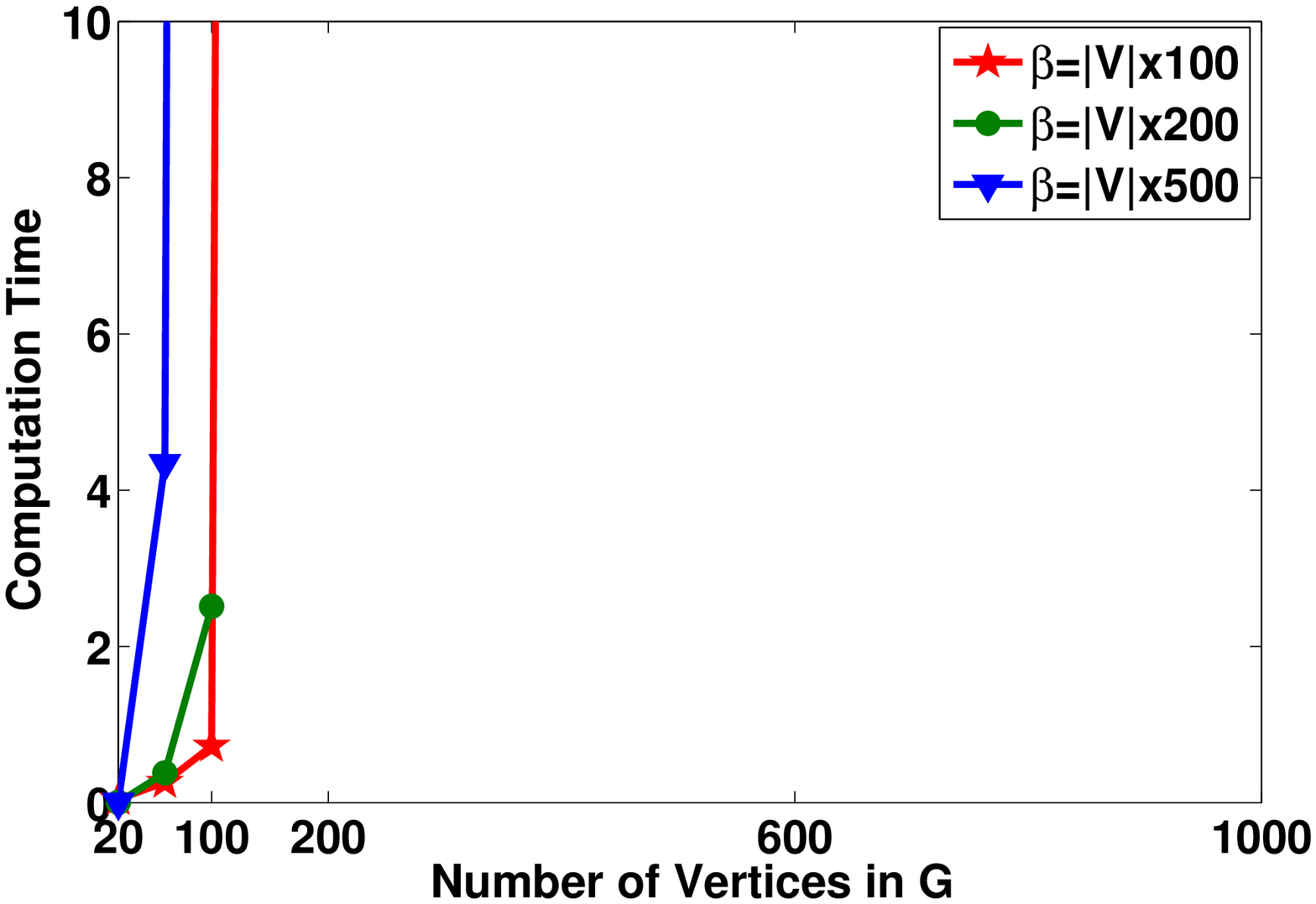}}
\subfigure[Scale-free graphs with
$\alpha=2.5$]{\label{fig:scalefree2}
\includegraphics [height=1.25in, width=.23\textwidth]{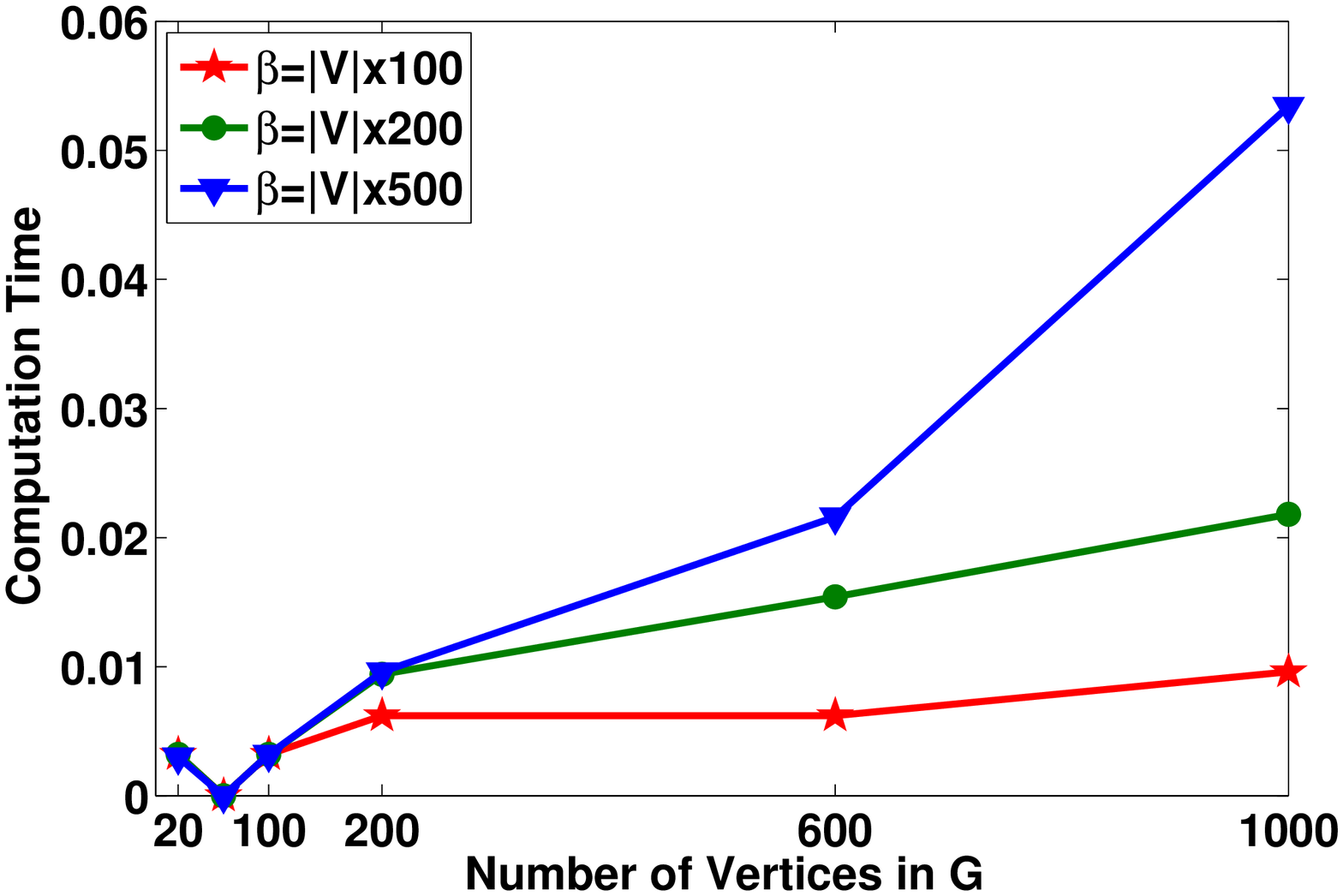}}
\caption{Average computation time of {\tt CPLEX} on problem classes
as a function of the number of vertices in G and the parameters of
the problem classes}
  \label{fig:opt}
\end{center}
\end{figure}

Figure \ref{fig:opt} shows how the optimal solution time of {\tt
CPLEX}, an exact method, varies depending on the problem size,
class, and structure. The $x$-axis and the $y$-axis represent the
number of vertices and the average computation time, respectively.
The right-most data point on a line shows the size of the largest
instance that can be solved to optimality in a group. In general,
its performance is good for small to medium scale graphs. However,
in our study, 39 out of 90, 73 out of 285 and 39 out of 180
instances in problem classes (a), (e) and (f), respectively, could
not be solved optimally using {\tt CPLEX} within the time limit.
This observation opens the door for heuristics to find acceptable
but possibly suboptimal solutions.

\paragraph{Random graphs} From figure \ref{fig:randopt} we conclude that for randomly connected
 graphs with more than 200 nodes, optimal solution is not achievable within the bounded time.
 It also suggests that the density of graphs is
a factor that affects the solvability. The solver does increasingly
better as the density $\eta$ goes down (up to 0.01) for the same
number of vertices. Its sensitivity with respect to the size and
density is apparent in the plots for $\eta$ equal to 0.05 and 0.10,
i.e., a 16 fold increase in solution time.

\paragraph{Bounded valance graphs} Compared to random graphs, figure \ref{fig:boundopt}
shows an improved performance on bounded valence graphs solving all
instances under 0.3 seconds. The reason for the performance
difference may be due to the considerably higher number of edges
 in a randomly connected graph (which forces the number constraints in the IP model to go higher)
than that of a bounded valence graph. However, although we expect
higher solution time for graphs with larger valence, figure
\ref{fig:boundopt} shows substantially higher time for valence 3
than valences 6 and 9 suggesting other factors may also be playing a
role.

\paragraph{Irregular bounded valance graphs} Although the degree distribution is neither
constant nor fully randomly distributed, CPLEX performs similarly to
bounded valence graphs. As figure \ref{fig:irboundopt} shows, all
solutions are computed in less than .25 seconds, and that the
computation time increases with the increase in valence.

\paragraph{Regular mesh graphs} Figure \ref{fig:meshopt} shows that the size of meshes (2D, 3D or 4D) usually does not have any influence on the performance barring the abrupt behavior of the 4D mesh graph. In general, the
solution time appears to linearly increase with the increase in
graph size, though the increase in time is extremely small.

\paragraph{Irregular mesh graphs} Unlike the irregular bounded valence graphs, mesh graphs are more susceptible to irregularity and the computation time substantially increases with the degree of irregularity. Figures \ref{fig:irmeshopt2} through \ref{fig:irmeshopt4} show that the
sizes of the problems that can be solved to optimality decrease and
the computation times increase with increasing degree of
irregularity. This result is quite reasonable and expected because
increasing irregularity increases the number of edges, and thus the
computation time as well. This is also because randomly adding edges
to a mesh graph makes it structurally more similar to random graphs,
which, as discussed earlier, is inherently hard to solve.

\paragraph{Scale-free graphs} We consider the
effect of the two parameters $\alpha$ and $\beta$ on the solvability
of the problems. On one hand, increasing $\alpha$ makes the degree
distribution sharper, i.e, we observe smaller number of vertices
with high degrees. On the other hand, increasing the value of
$\beta$ increases the degrees of non-hub nodes. Figures
\ref{fig:scalefree1} and \ref{fig:scalefree2} represent the optimal
solution times of scale-free instances. It is clear that the
difficulty of the problem is closely related to parameters $\alpha$
and $\beta$. The figures show that computation times decrease
significantly with increasing values of $\alpha$. When $\alpha=1.5$,
the instances with more than 100 vertices cannot be solved to
optimality within the time limit. When $\alpha=2.5$, however, all of
the instances can be solved optimally in less than 0.06 seconds.
These figures also show that the computation time increases with
increasing values of $\beta$. This means that increasing degrees of
non-hub nodes makes the problem more difficult.


\subsection{Performance Profile of Solution Methods}

\begin{figure}[t!]
\begin{center}
\subfigure[Randomly connected graphs]{\label{fig:perfrand}
\includegraphics [height=1.25in, width=.23\textwidth]{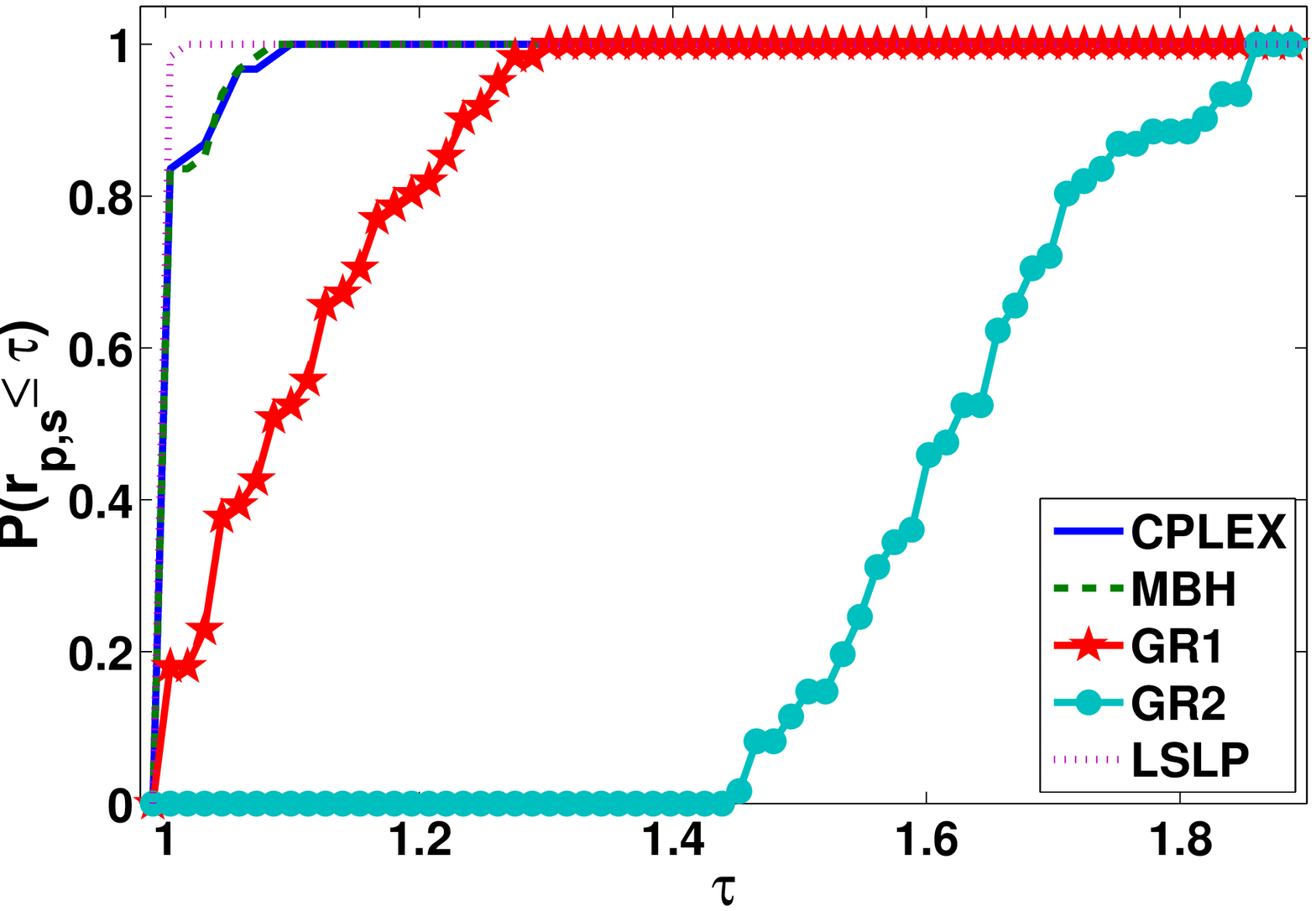}}
\subfigure[Bounded valence graphs]{\label{fig:perfbound}
\includegraphics [height=1.25in, width=.23\textwidth]{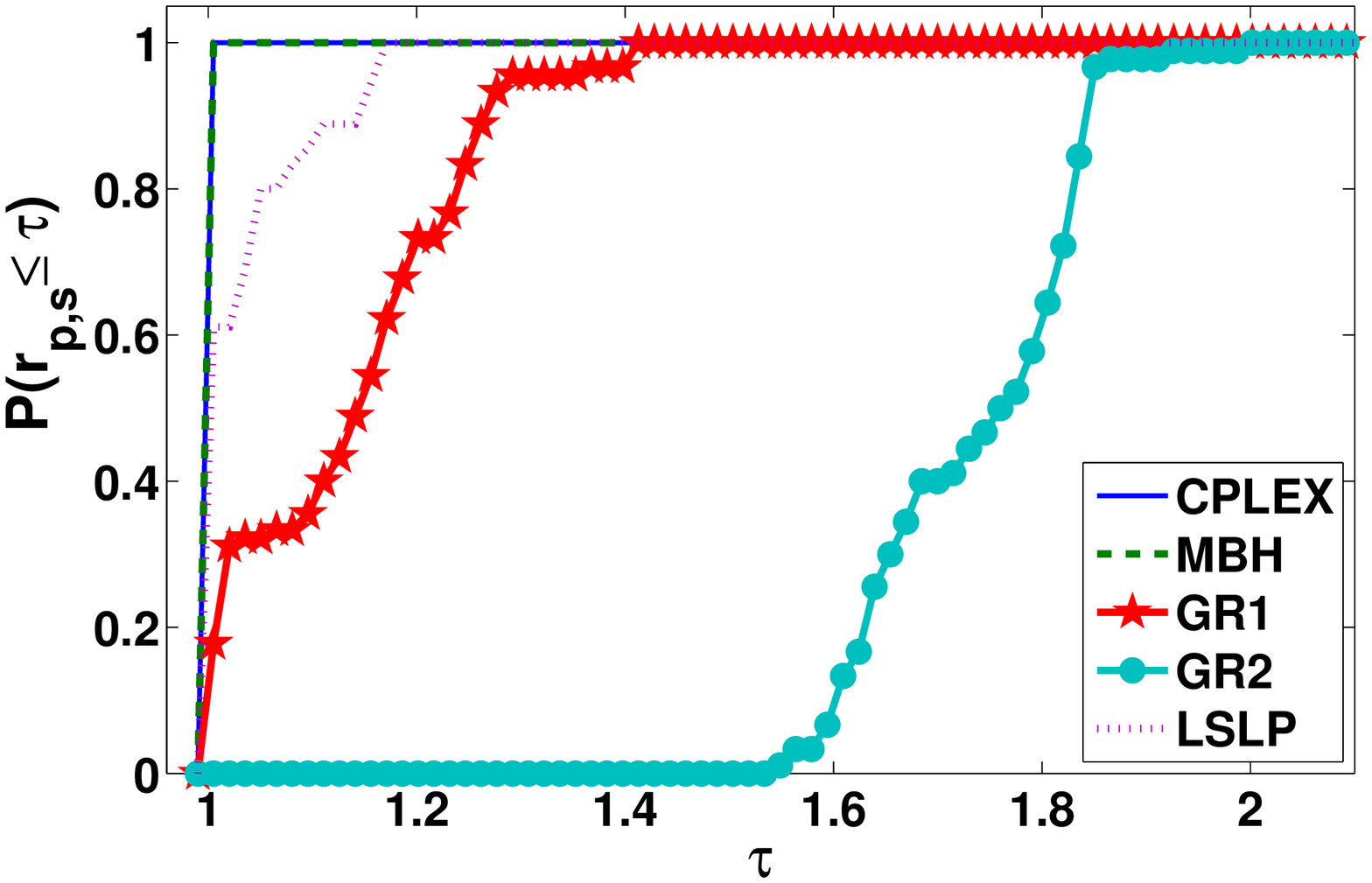}}
\subfigure[Irregular bounded valence graphs]{\label{fig:perfirbound}
\includegraphics [height=1.25in, width=.23\textwidth]{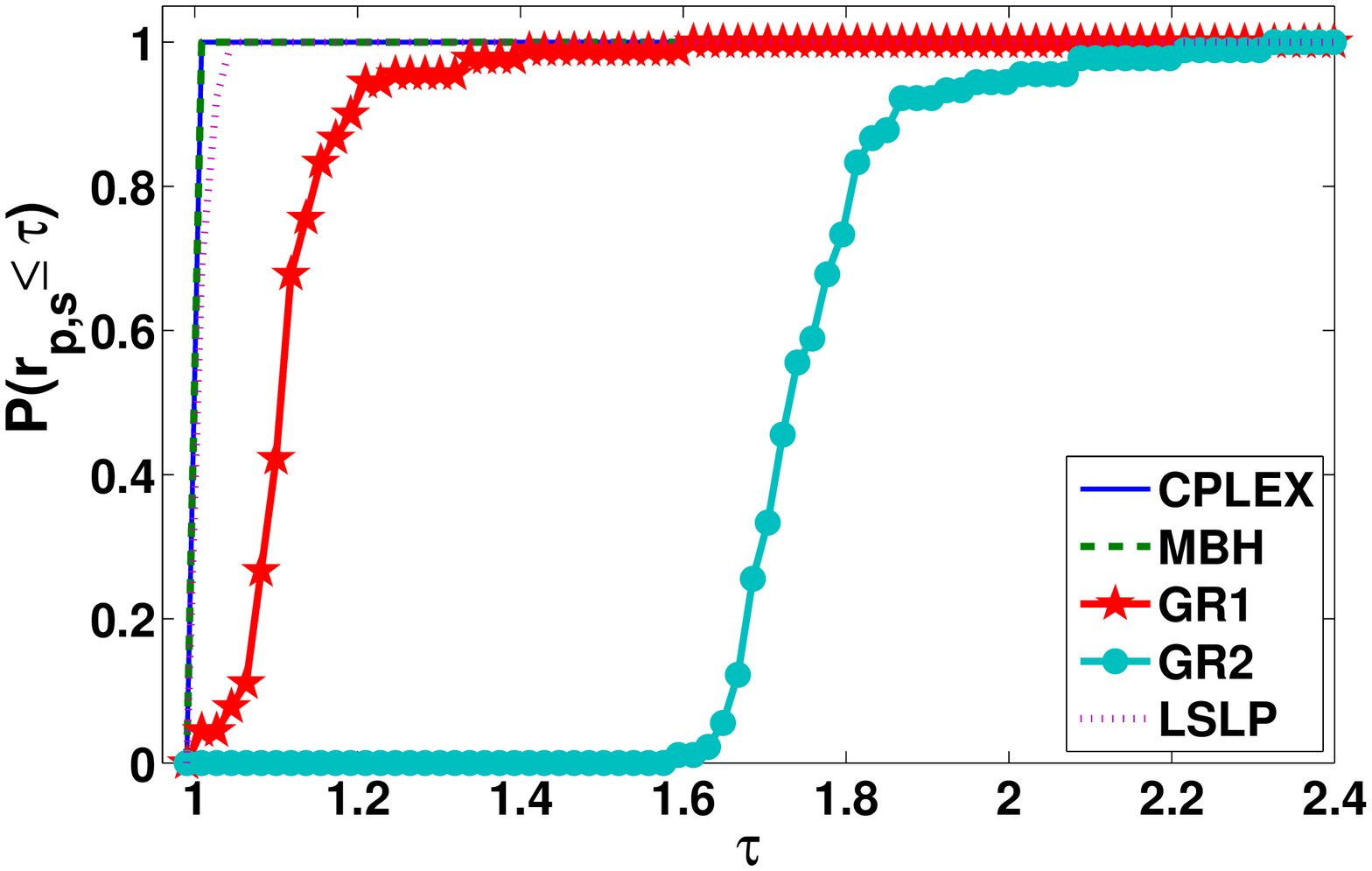}}
\subfigure[Regular meshes with 2D, 3D and 4D]{\label{fig:perfmesh}
\includegraphics [height=1.25in, width=.23\textwidth]{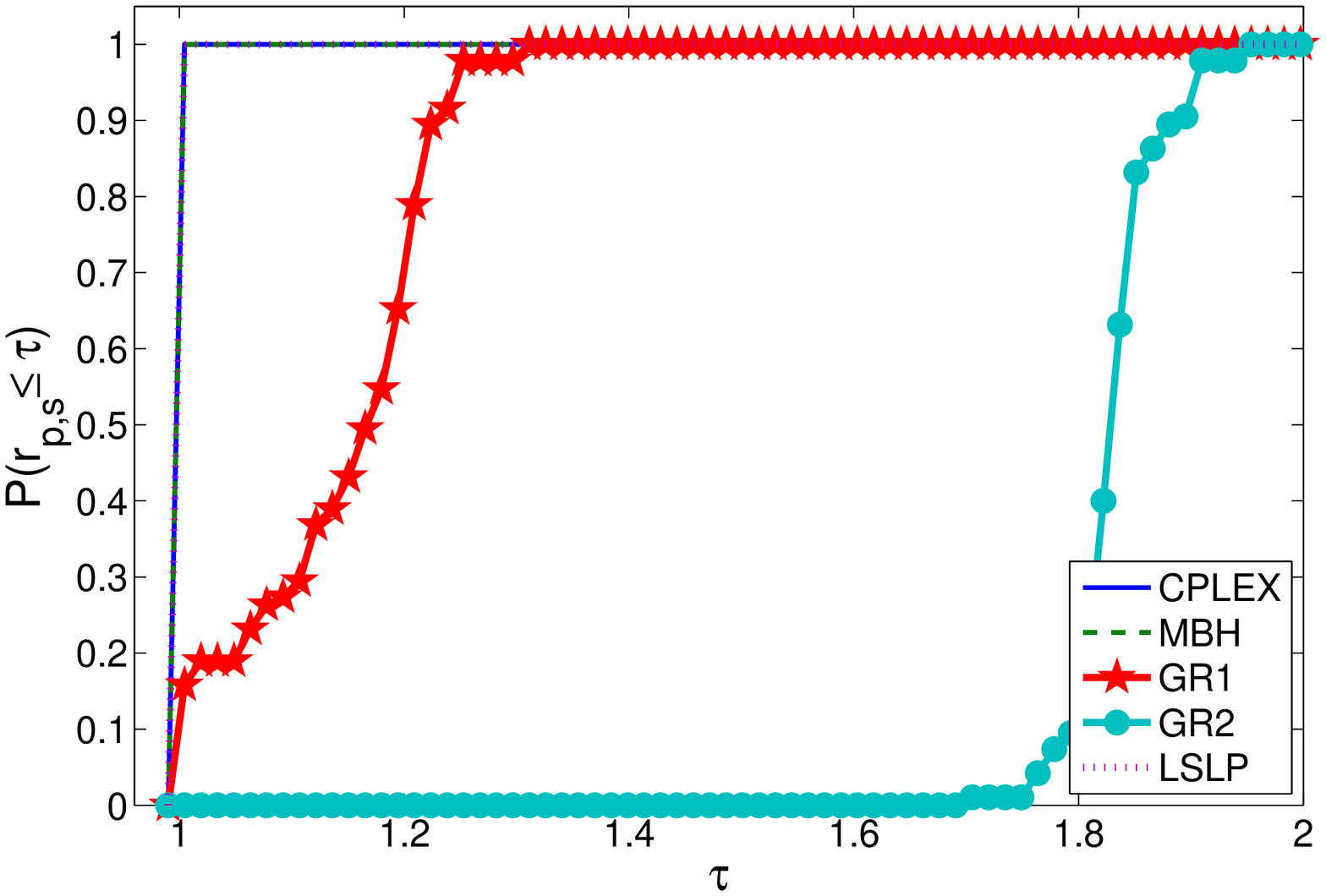}}
\subfigure[Irregular meshes]{\label{fig:perfirmesh}
\includegraphics [height=1.25in, width=.23\textwidth]{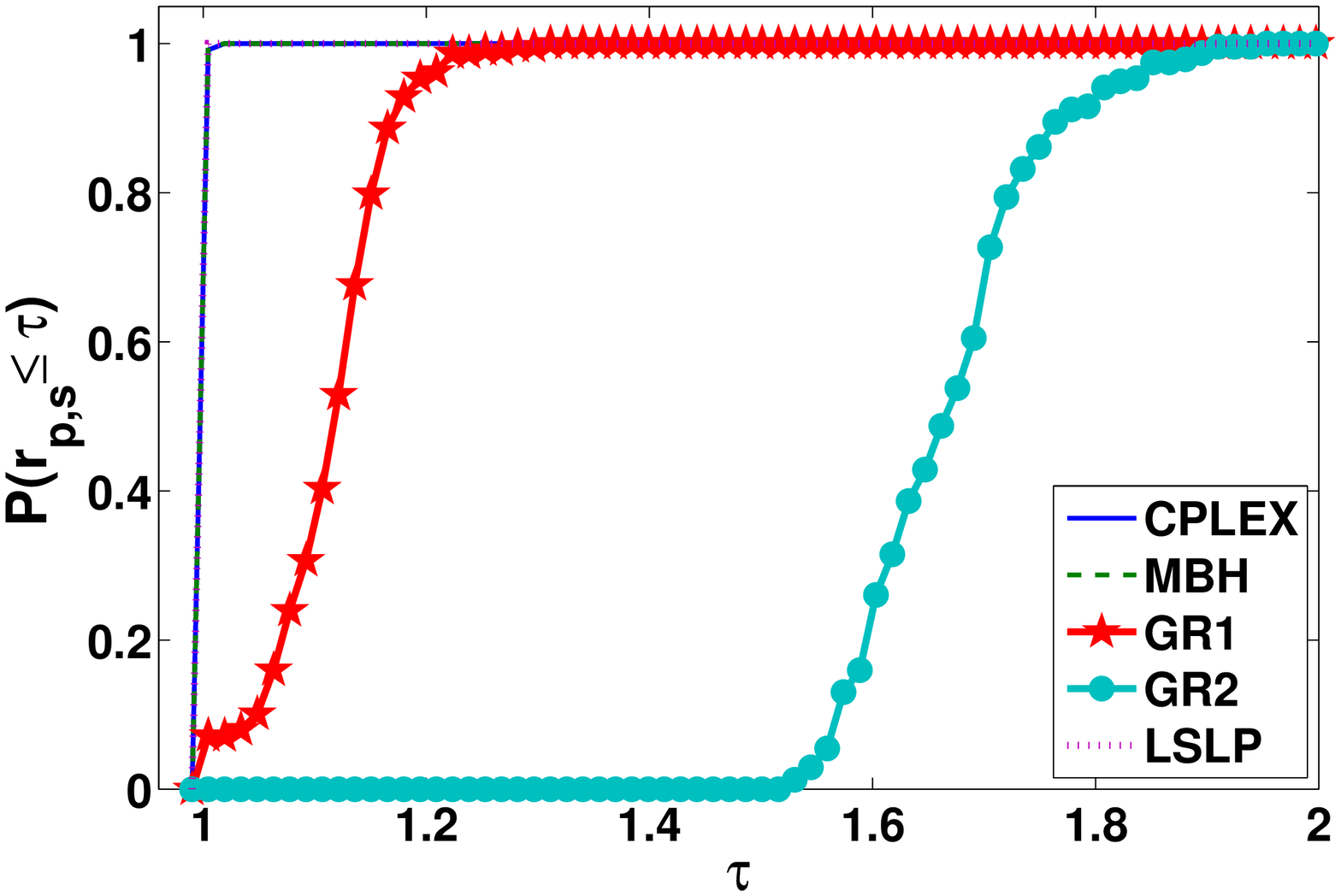}}
\subfigure[Scale-free graphs]{\label{fig:perfscale}
\includegraphics [height=1.25in, width=.23\textwidth]{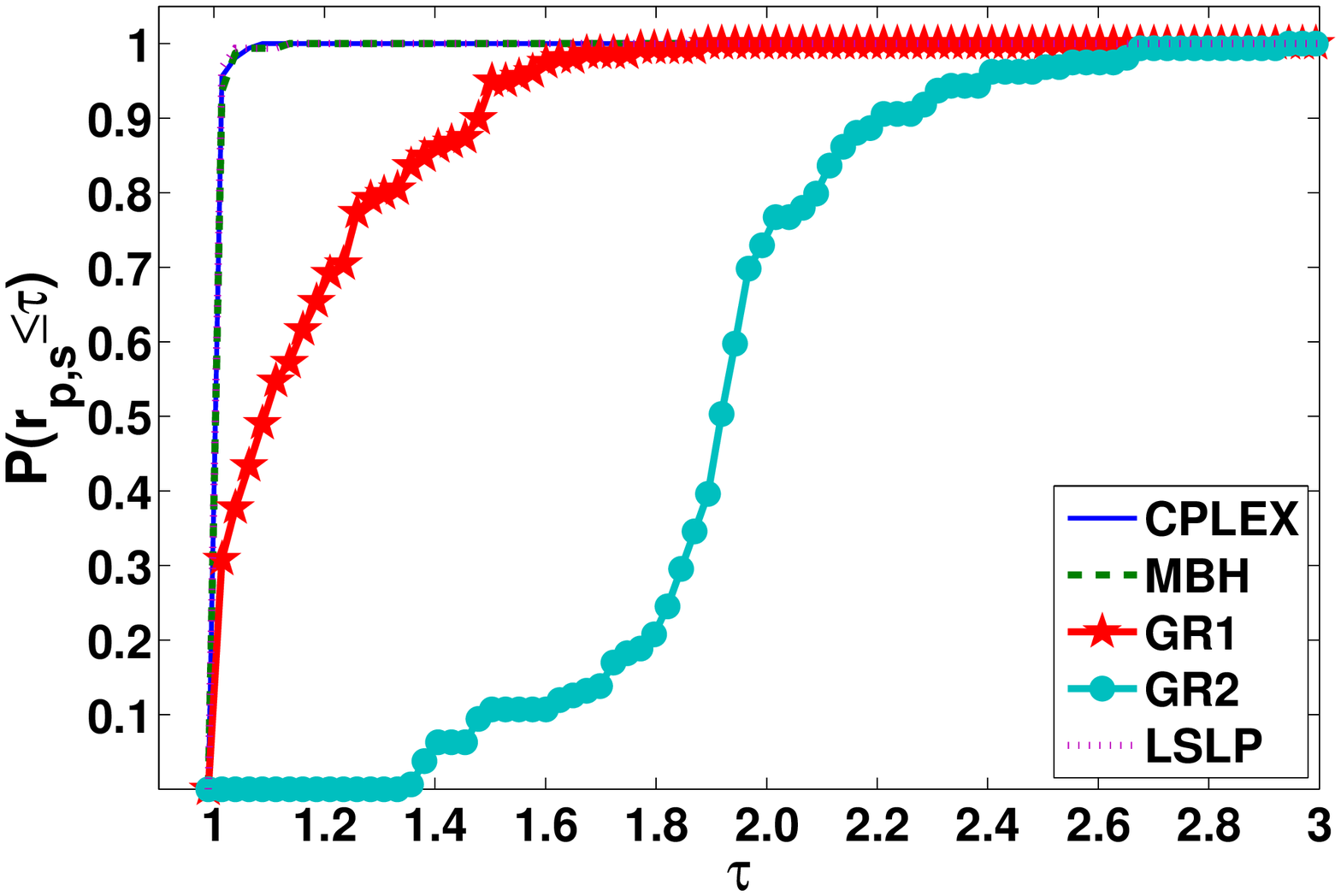}}
\caption{Performance profiles for the algorithms on the problem
classes in terms of solution quality}
  \label{fig:perfpro}
\end{center}
\end{figure}

To study the quality of solutions generated by other solution
methods with respect to the optimal solutions computed using CPLEX,
we refer to figures \ref{fig:perfrand} through \ref{fig:perfirmesh}.
These plots are called performance profiles of algorithms that
depict the fraction of problems for which the algorithm is within a
factor of the best solution \cite{Dolan02}. Thus, they compare the
performance of an algorithm $s$ on an instance $p$ with the best
performance observed by any other algorithm on the same instance.
The x-axis represents the \textit{performance ratio} given by
\[
r_{p,s}=\frac{\alpha_{p,s}}{\min\{\alpha_{p,s}:s\in S\}},
\]
where $\alpha_{p,s}$ is the number of hub nodes in the hub cover
when the instance $p$ is solved by algorithm $s$ and $S$ is the set
of all benchmark algorithms. The y-axis shows the percentage of the
instances that gives a solution that is less than or equal to $\tau$
times the best solution. Recall that {\tt CPLEX} cannot solve all of
the instances in problem classes (a), (e) and (f) to optimality.
However, the solver is able to find feasible solutions for some of
those unsolved instances (11 out of 39, 44 out of 73, and 24 out of
39 in (a), (e), and (f) respectively). Figure \ref{fig:perfpro}
includes all instances except those for which {\tt CPLEX} cannot
find either feasible or optimal solutions within the time limit.

We first analyze how much we sacrifice from the optimality by
employing mathematical programming-based heuristics MBH and LSLP.
Recall that MBH and LSLP solve the same restricted problem. While
MBH tries to solve the problem to optimality, the latter visits
alternate solutions in the feasible region. Figure
\ref{fig:perfrand} shows that 12\% of the instances in class (a)
where both MBH and LSLP find better feasible solutions than that of
{\tt CPLEX}. Note that, this can happen if and only if {\tt
  CPLEX} returns a feasible solution rather than an optimal solution
within the time limit. For other problem classes, the performances
of the {\tt CPLEX} and MBH are quite similar. Moreover, these
figures show that LSLP is outperformed by MBH and {\tt CPLEX} on
almost 40\% and 30\% of instances in problem classes (b) and (c)
respectively. For the remaining problem classes, the performance of
LSLP is also comparable to MBH and {\tt CPLEX}.


The greedy algorithms return feasible but sub-optimal solutions
quickly. Except the scale-free networks, the performances of the
greedy algorithms do not change with respect to problem classes. GR2
is known as 2-approximation algorithm for MVC. Figures
\ref{fig:perfirbound} and \ref{fig:perfscale} show that there are
some instances for which performance ratios of GR2 are higher than
2. Obviously, the approximation ratio of GR1 for MHC problem is
higher than 2. Intuitively, the performance of GR1 is supposed to be
better when the degree distribution of the vertices is not uniform.
Since the average degrees of the vertices are identical or are quite
similar for the instances in problem classes (a)-(e), the
performance of GR1 does not vary for these problem classes. However,
figure \ref{fig:perfscale} shows that GR1 finds the optimal or the
best solution in 30\% of the scale-free instances. This means that
the performance of the GR1 is better for the graphs that follow the
power-law distribution.

\subsection{Cost Profile of Solution Methods}

\begin{figure}[h]
\begin{center}
\subfigure[Randomly connected graphs]{\label{fig:timerand}
\includegraphics [height=1.25in, width=.23\textwidth]{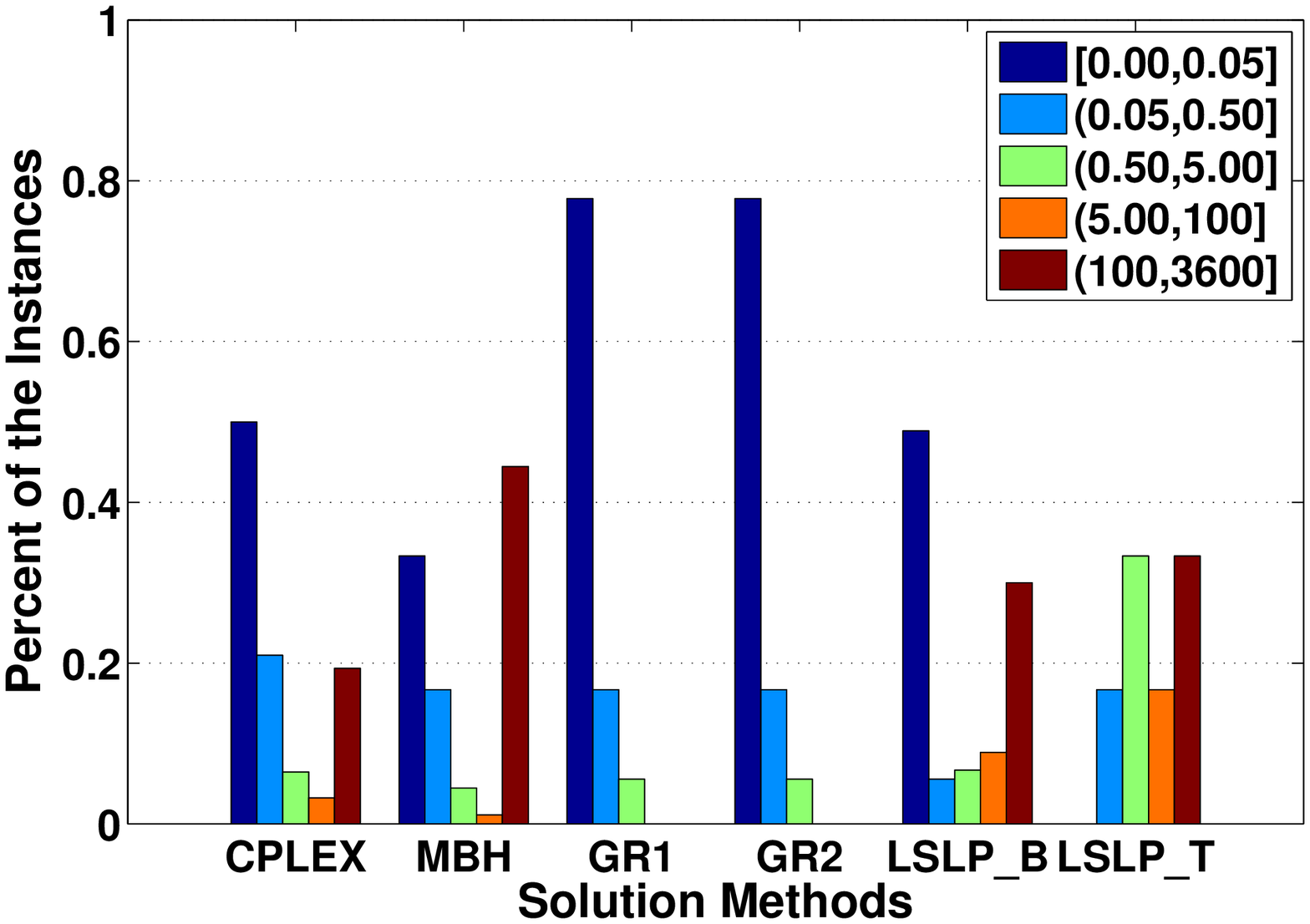}}
\subfigure[Bounded valence graphs]{\label{fig:timebound}
\includegraphics [height=1.25in, width=.23\textwidth]{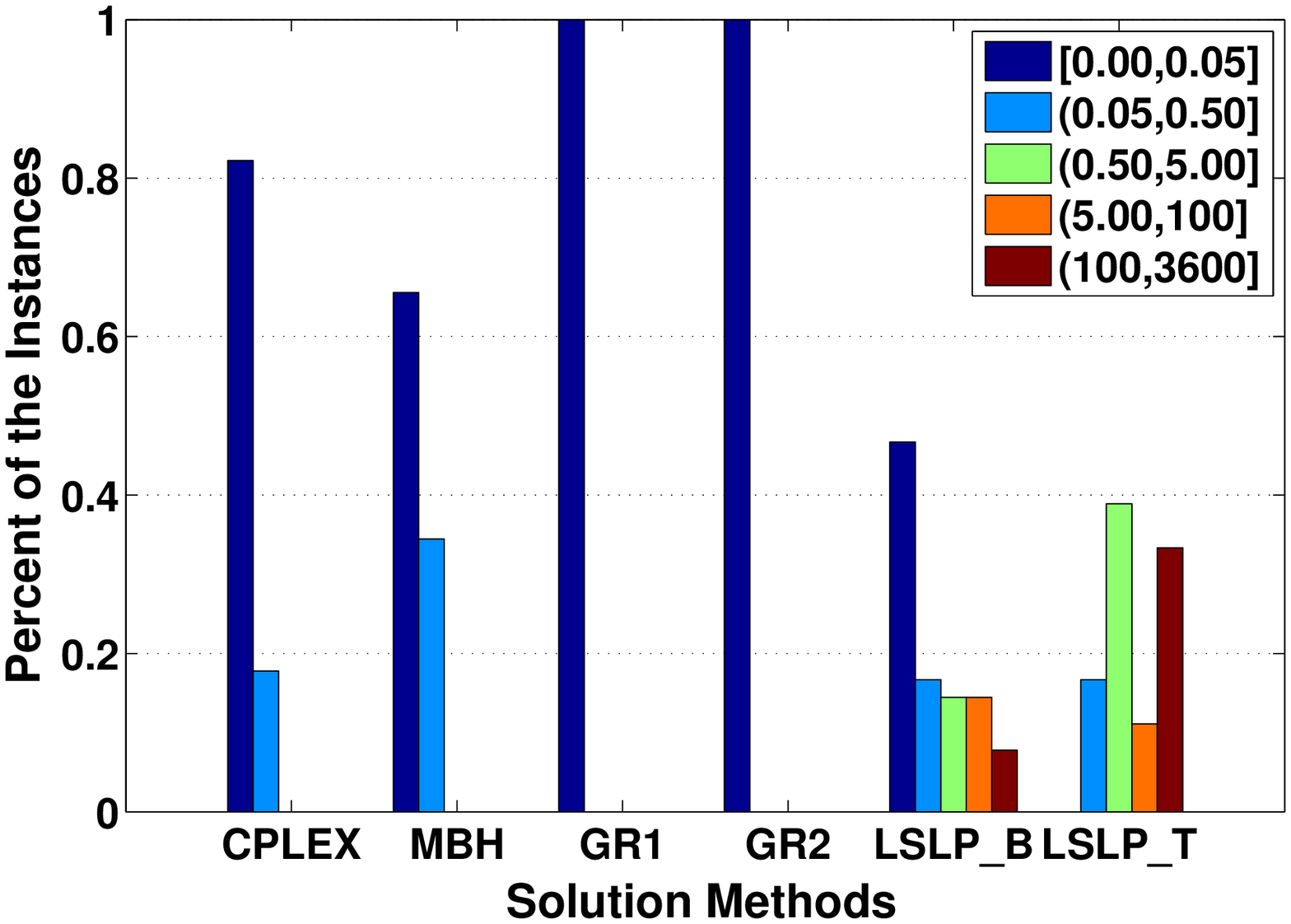}}
\subfigure[Irregular bounded valence graphs]{\label{fig:timeirbound}
\includegraphics [height=1.25in, width=.23\textwidth]{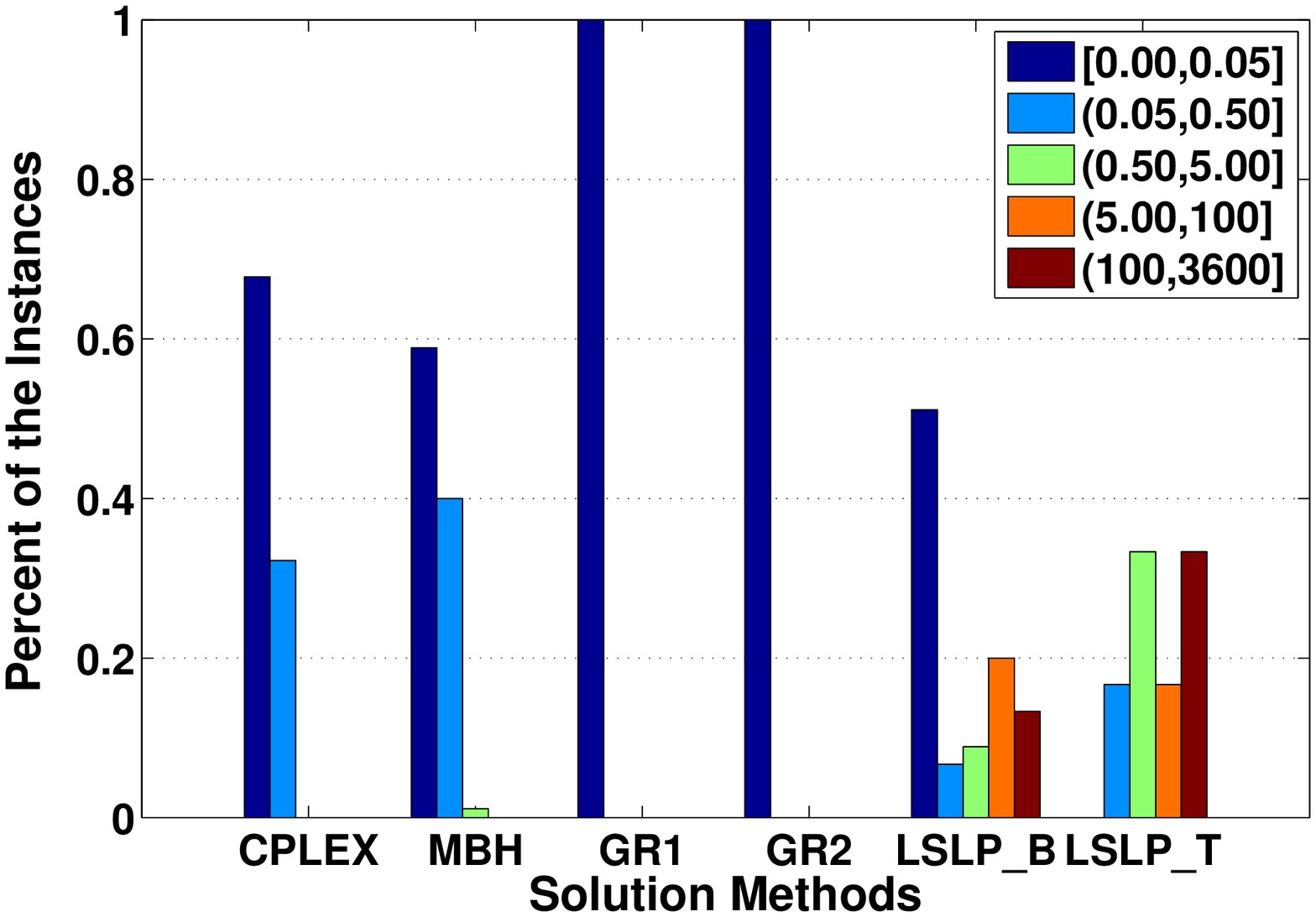}}
\subfigure[Regular meshes with 2D, 3D and 4D]{\label{fig:timemesh}
\includegraphics [height=1.25in, width=.23\textwidth]{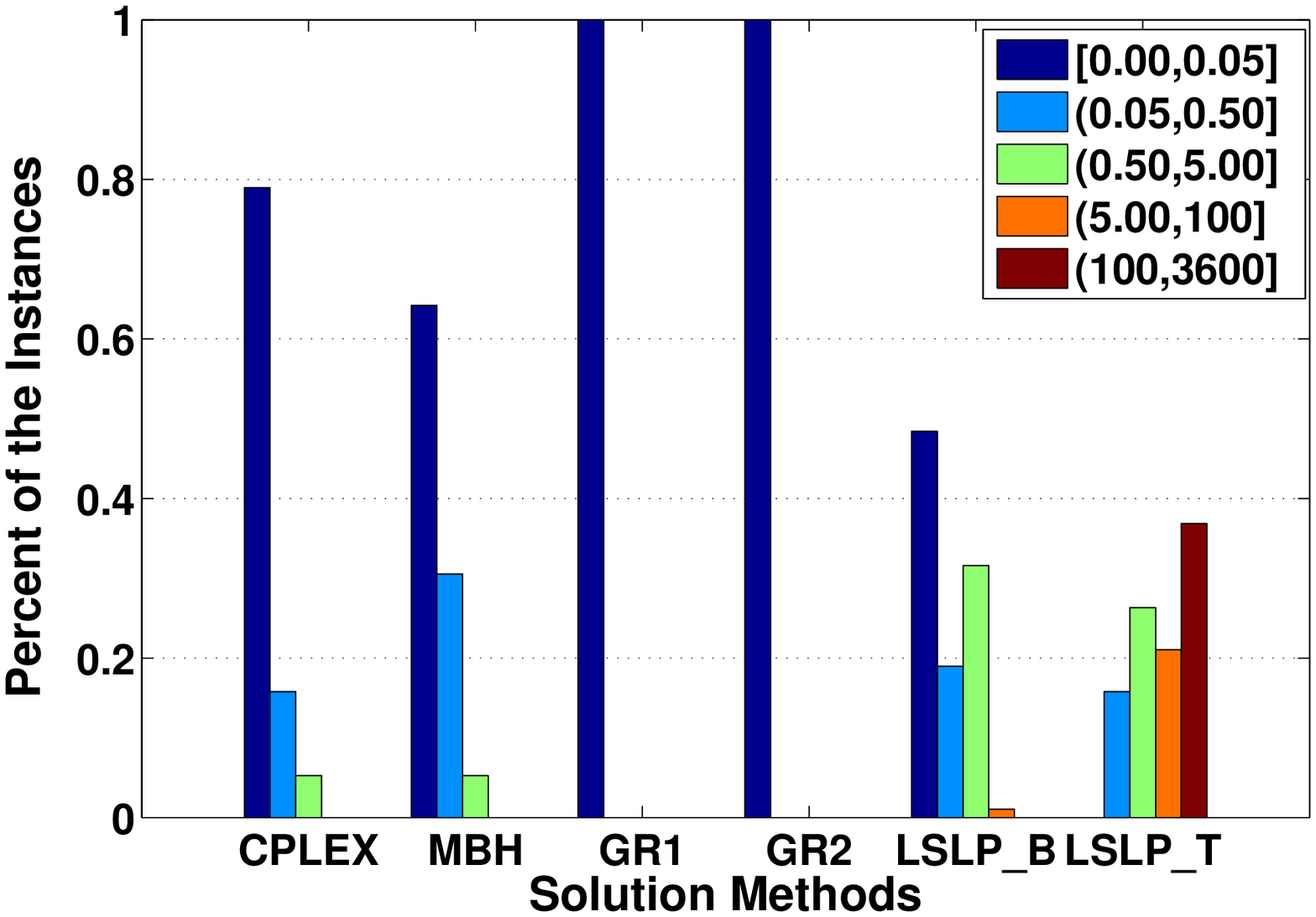}}
\subfigure[Irregular meshes]{\label{fig:timeirmesh}
\includegraphics [height=1.25in, width=.23\textwidth]{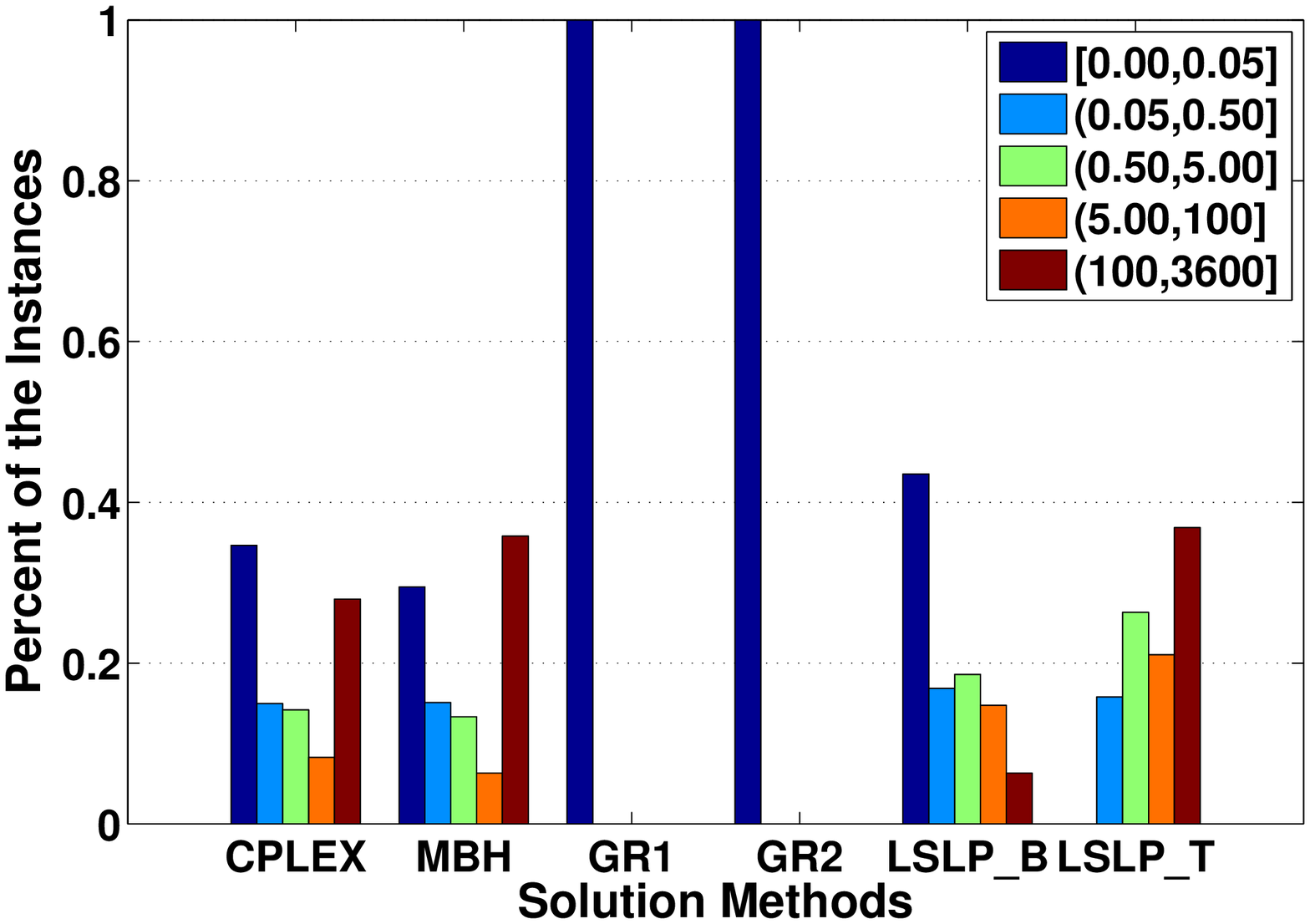}}
\subfigure[Scale-free graphs]{\label{fig:timescale}
\includegraphics [height=1.25in, width=.23\textwidth]{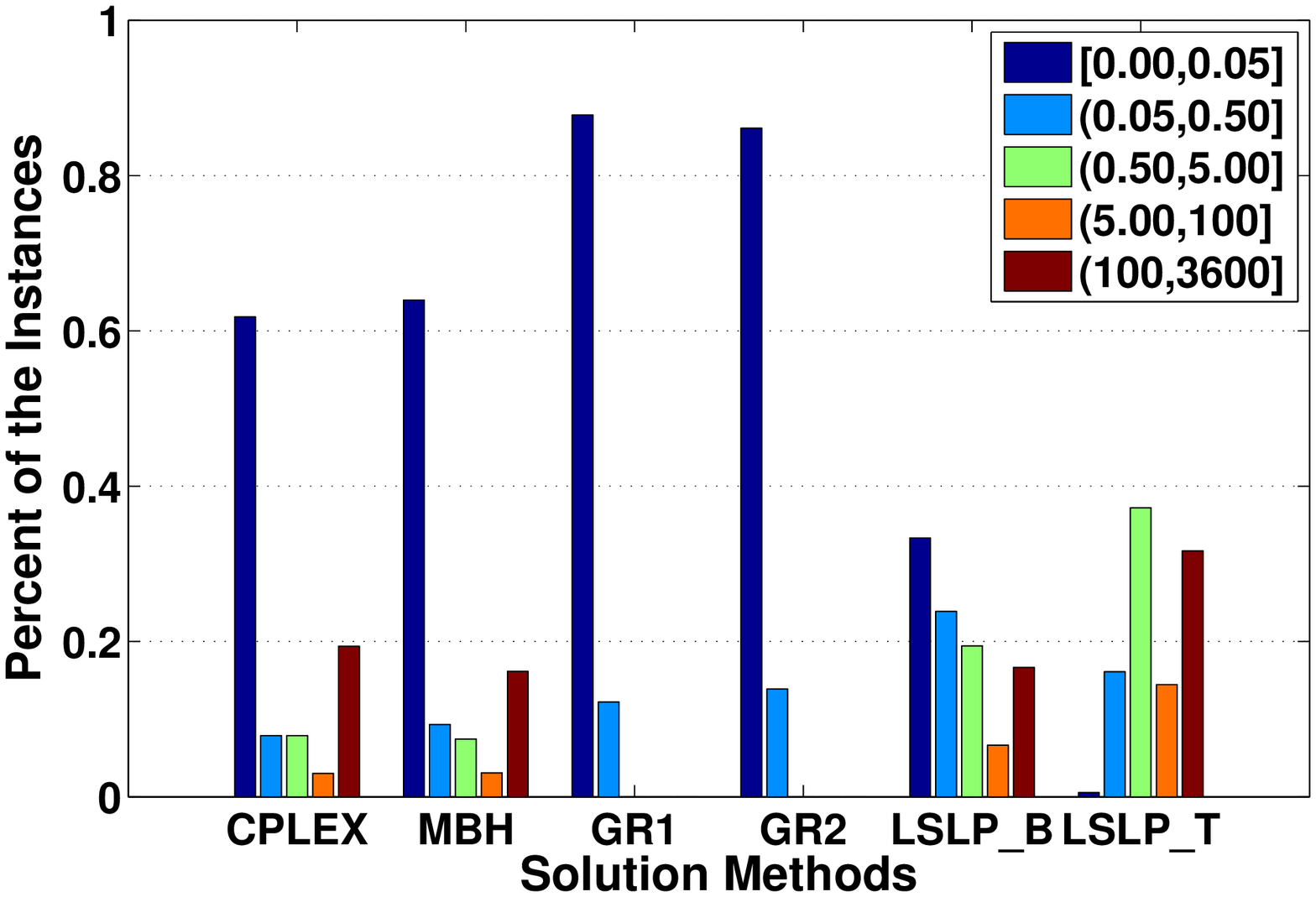}}
\caption{Computation time distributions of the solution methods on
the problem classes}
  \label{fig:timedist}
\end{center}
\end{figure}

The previous two analyses focused on the optimal solvability and the
quality of the solutions. In this section we turn our attention to
the cost of computing a feasible or optimal MHC solutions in terms
of time. Figures \ref{fig:timerand} through \ref{fig:timescale}
summarize the distribution of the average computation times of the
algorithms over the problem classes. In these plots, the instances
for which feasible solutions were not found by any of the algorithm
within a time limit are excluded. Each bar in the figure represents
the percentage of the instances that are solved within the time
interval stated in the legend, e.g., the blue bar for 0.0 to 0.05
seconds. Since LSLP is a local search algorithm, we show both the
total computation time and the first time when the best solution is
found.

The results clearly show that the solution times of greedy
algorithms (GR1, GR2) are much shorter than that of the other
algorithms. MBH can solve the restricted problem to optimality in a
reasonable amount of time for a great majority of the instances. We
have already discussed earlier that the performance of the MBH is
good in terms of its solution quality. However, the main drawback
for MBH is its inability to solve the restricted problem to
optimality. In such cases, LSLP may serve as an alternative to MBH
as it is comparable to MBH in terms of both solution quality and
time, and because LSLP is a local search algorithm, it is also
guaranteed to produce a feasible solution. However, the performance
of LSLP is dependent upon prudent selection of algorithmic
parameters, e.g., the total number of iterations, the number of
improvement iterations (see \cite{Yelbay12} for details). There is a
trade-off between solution time and the solution quality. Decreasing
the total number of iterations may result in a decrease in the total
solution time. However, it may increase the optimality gap.

\section{Future Research Opportunities}
\label{sec:fut}

Constraint solvers such as CPLEX usually do not offer all optimal
solutions. Such solutions also do not exploit database meta-data in
computing the most desired solution. For example, for the graph in
figure \ref{fig:example}(c), there is no guarantee that the solver
will produce the desirable solution $\{w_5,w_4\}$ when the data
graph is known to be $d$. Therefore, we are required to compute all
possible MHCs of $q_2$, i.e., $\Gamma(q_2)$, so that we are able to
identify this least cost query plan. Unfortunately, it is not
guaranteed that solvers can even always compute a solution, let
alone the whole family of solutions $\Gamma(Q)$ of $Q$.

The discussion in section \ref{subsec:eval} also suggests that
although the cost of computing MHC is significantly low for small
graphs, it still remains high for many graph types when the query
graph is large. Therefore, even though some of the existing solvers
may be useful for applications involving small scale-free graphs
such as protein-protein interaction networks, they may not be a
great candidate for big data applications in social networks and
world wide web. Though it may be challenging, we believe the low MHC
solution time for many graph types offers hope that designing
algorithms for $\Gamma(Q)$ is feasible for most practical
applications, but remains as an interesting problem.

It is also important to recognize that while developing the least
cost MHC using meta-data may be feasible for a single data graph,
devising such algorithm for a large set of data graphs may not be
feasible. It is thus worth investigating if a general but a single
optimal query plan (without computing $\Gamma(Q)$), for which we
have a solution, can be dynamically adjusted for best performance
over a set of graphs. Finally, it remains an open question if a
suboptimal MHC produced by a greedy algorithm can be improved enough
to defeat or match the overall processing performance using an
optimal MHC solution, i.e., total cost of MHC, plan generation, plan
selection and execution.

\section{Conclusions}
\label{sec:conl}

In this paper we have formally introduced the idea of graphlets as a
basic unit for graph representation in a way similar to RDF triple
store, and the concept of minimum hub cover of graphlets as a basic
ingredient toward graph query optimization. We have demonstrated on
intuitive grounds that such an approach can leverage generic access
structures such as hash \cite{Chung92} and set indices
\cite{TerrovitisBVSM11,Mamoulis03} for query optimization. Though
computationally hard, we have also demonstrated that query
processing and optimization using MHC and subgraph isomorphism is
computationally feasible and intellectually intriguing. In
particular, we have shown that for many application domains of
current interest such as social networks, and protein-protein
interaction networks, existing constraint solvers are capable of
delivering optimal solutions for MHC, and therefore can be used to
develop optimization strategies. It is our thesis that covering
based graph processing we have presented opens up new research
directions and holds enormous promise. The logical next step is to
develop a query processor by integrating the algorithms in sections
\ref{sec:cost} and \ref{sec:qp-alg}, with new algorithms outlined in
section \ref{sec:fut}. These are some of the tasks we plan to
continue as our future research.

\section{Acknowledgements}
This study was supported partially by TUBITAK 2214 Ph.D. Research
Scholarship Program.
\bibliographystyle{IEEEtran}

\end{document}